\documentclass[10pt, conference, compsocconf]{IEEEtran}

\usepackage{booktabs}
\usepackage{tikz}
\usepackage{url}
\usepackage{pdfpages}
\usepackage{afterpage}
\usepackage{color, colortbl}
\usepackage{graphicx} 
\usepackage{amsmath}
\usepackage{subfigure}
\usepackage{footmisc}
\usepackage{epstopdf}
\usepackage{caption}
\ifCLASSINFOpdf
\else
\fi
\begin{document}
%
\title{GPGPU Performance Estimation with Core and Memory Frequency Scaling}


\author{\IEEEauthorblockN{Qiang Wang}
\IEEEauthorblockA{Department of Computer Science\\
	Hong Kong Baptist University, Hong Kong\\
	qiangwang@comp.hkbu.edu.hk}
\and
\IEEEauthorblockN{Xiaowen Chu}
\IEEEauthorblockA{Department of Computer Science\\
	Hong Kong Baptist University, Hong Kong\\
	HKBU Institute of Research and Continuing Education,\\
	Shenzhen, China\\
	chxw@comp.hkbu.edu.hk}
}



%


\maketitle

\begin{abstract}
Graphics Processing Units (GPUs) support dynamic voltage and frequency scaling (DVFS) in order to balance computational performance and energy consumption. However there still lacks simple and accurate performance estimation of a given GPU kernel under different frequency settings on real hardware, which is important to decide best frequency configuration for energy saving. This paper reveals a fine-grained model to estimate the execution time of GPU kernels with both core and memory frequency scaling. Over a 2.5x range of both core and memory frequencies among 12 GPU kernels, our model achieves accurate results (within 3.5\%) on real hardware. Compared with the cycle-level simulators, our model only needs some simple micro-benchmark to extract a set of hardware parameters and performance counters of the kernels to produce this high accuracy. 

\end{abstract}


%
\IEEEpeerreviewmaketitle

\section{Introduction}

Recently the Graphics Processing Units (GPUs) are becoming widely used from Deep Learning (DL) workstations to high performance supercomputing centers. 
In particular, most popular DL toolkits \cite{collobert2011torch7,abadi2015tensorflow,jia2014caffe,cntkspeedcomparasion2016} heavily rely on the remarkable computation power of GPUs. However, due to rapidly increasing computation requirements of both DL toolkits and other GPUs applications on large amount of data, the total energy consumption can be very high, which not only results in high electricity budgets but also violates green computing. 
For example, the supercomputer Titan accelerated with the NVIDIA K20x requires a power supply of 8.21 million Watts with an electricity cost of about 23 million dollars per year \cite{TitanIntro}. Even decreasing 5\% of the power consumption can reduce up to 1 million dollars of electricity costs. Effective energy saving techniques are emergent to be designed for GPUs.

Energy conservation techniques on modern computers are generally based on Dynamic Voltage and Frequency Scaling (DVFS). Nowadays GPUs usually support simple automatic voltage and frequency adjustment in order to save power and protect the hardware. Nevertheless, GPUs hardly gain the best energy efficiency under the default voltage and frequency settings \cite{mei2016survey, bridges2016understanding, greenmetrics2017power, hpca2018power} and still have potentials of energy conservation. To find the most energy efficient DVFS configurations, the energy consumption under different DVFS settings should be predicted, which requires modeling both performance and runtime power of GPUs under various settings of voltage and frequency. In this paper we would like to address the performance modeling problem. 

There are three main challenges of GPU performance prediction under different core and memory frequencies. First, compared to traditional CPU, GPUs have much more complex memory hierarchy of which GPU vendors reveal few details. Second, GPUs have two independent frequencies belonging to core and memory respectively and they affects different components of GPUs. Third, resource contention is heavy due to large number of concurrent threads. 

Some previous state-of-the-art work reveals analytical pipe-line GPU performance model \cite{hong2009analytical, hong2010integrated, nath2015crisp, miftakhutdinov2012predicting,sim2012performance,chen2014run} which emphasizes the relationship between compute cycles and memory latency. However, there still exist some opportunities to reinforce the model. First, the L2 cache becomes larger and larger among the evolutionary GPU generations. Compared to Fermi 2011, Maxwell 2014 has four times larger L2 cache. Larger cache generally can increase the cache hit rate, which reminds us to consider more on L2 cache latency and throughput instead of DRAM. Second, most of the previous models only work under the default frequency settings of GPU. The kernel behavior could change significantly when the core and memory frequencies have been adjusted.  

Simulation methods \cite{leng2013gpuwattch, lucas2013single, bakhoda2009analyzing} expose sufficient details to help understanding the execution of GPU kernels. The best available simulator to date \cite{aamodt2012gpgpu} combines performance counters and specific hardware information to estimate the kernel execution time with high accuracy. However, compared with the fast evolving GPU generations, such simulators still stand for the earlier GPU architecture like Fermi, which does not meet the great changes of newer hardware. Besides, these simulators usually consume much longer time than real hardware, which are difficult to be applied to real-time power-performance optimization. 

Recent GPU performance models \cite{wu2015gpgpu, abe2014power, song2013simplified, nagasaka2010statistical,dao2015performance} also witness the trend of Machine Learning methods such as K-means, multiple linear regression and neural network. and obtain considerable accuracy. However, few of them introduce frequency scaling as impact factors in their models. Besides, their works strongly rely on training data such as specific performance counters and kernel settings. Even they can reveal some correlations between the input parameters and execution time, it needs further explorations of how they interact with each other and contribute to the final time prediction.

We believe that a fast and accurate GPU performance model is a key ingredient for energy conservation with DVFS technique and it should be applicable to real hardware. In this paper, we first attempt to model the memory system of GPU with a FCFS (First-come-first-serve) queue in which service rate depends on the memory frequency. Based on that, we propose a GPGPU performance estimation model considering both core and memory frequency scaling. Our paper reveals following contributions:
\begin{enumerate}
	\item We model the memory system of GPU with a simple queue related with the frequency. 
	\item We establish an analytical GPU performance model with both core and memory frequency scaling. 
	\item On a real GPU hardware, our performance model achieves 3.5\% MAPE (Mean Average Percentage Error) across 49 frequency settings with up to 2.5x scaling among 12 kernels. Meanwhile, we achieve 0.7\% to 6.9\% MAPE for each single kernel, which suggests great accuracy and low variance of our performance model. 
\end{enumerate}

The rest of this paper is organized as follows. Section \ref{sec:BM} introduces some basic knowledge about GPU and DVFS techniques followed by a motivating example about performance scaling behaviors with different frequency settings. Section \ref{sec:RW} lists some related work. Section \ref{sec:mem_model} details our memory queuing model for GPU with frequency scaling and based on it, Section \ref{sec:GPM} proposes our GPGPU performance estimation model with both core and memory frequency scaling. Section \ref{sec:Ex} describes our experimental setup and presents the experimental results. Finally we state our conclusion and future work in Section \ref{sec:cc}.

\section{Background and Motivation} \label{sec:BM}
\subsection{GPU Architecture}
\begin{figure}[ht]
	\centering
	\includegraphics[width=0.4\textwidth]{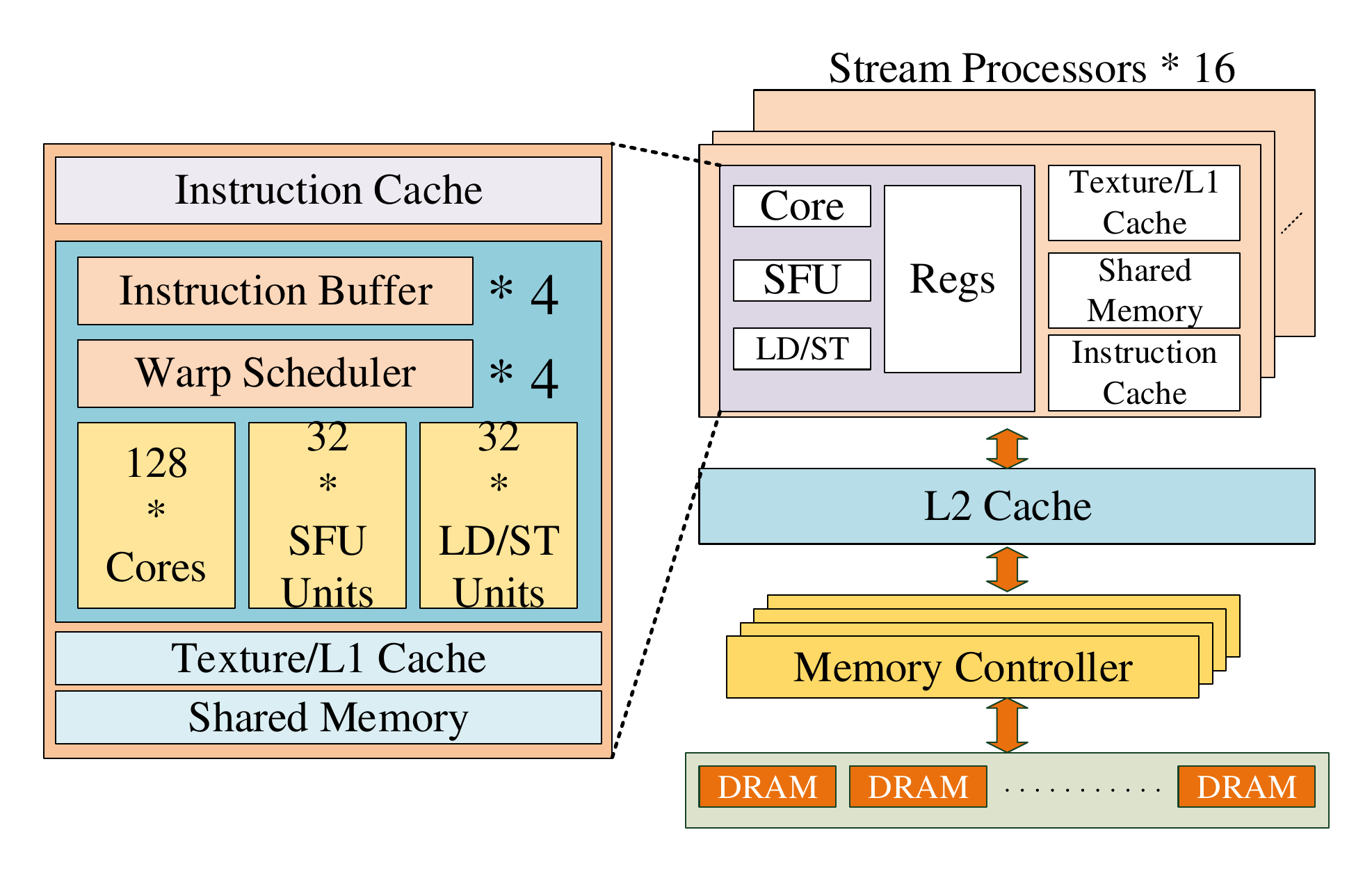}\\
	\caption{The block diagram of NVIDIA GTX980 GPU board.}\label{fig:arch_maxwell}
\end{figure}

Over past five years NVIDIA has released five generations of GPUs. The new functions and improvements through each updated version can be obtained from \cite{cudaProgram}. Despite of some different hardware configuration like core number and memory size, the basic chip framework is almost the same. Fig. \ref{fig:arch_maxwell} shows a brief block diagram of Maxwell GTX980 GPU. The structure details can be found in its official white paper. Note that GPUs have complicated memory hierarchy including dynamic random-access memory (DRAM), L2 cache, shared memory, texture/L1 cache and registers. Different memory types have their own characteristic in terms of latency, bandwidth and access scope, which makes it difficult to predict the execution time of a GPU kernel. 
\subsection{Dynamic Voltage and Frequency Scaling}
DVFS is one of the most typical techniques of energy conservation for traditional CPUs. The dynamic power is usually estimated by Equation \eqref{eq:dyn_power} \cite{kursun2006supply}. Since the total energy consumption of one application is obtained by multiplying the average runtime power and the execution time, performance modeling plays an important role in energy consumption prediction with different DVFS settings. For traditional CPUs, scaling up the frequency is usually a good option to save energy \cite{kim2015racing}. However, some previous GPU DVFS work indicates that GPUs have more complex energy scaling behaviors when adopting different voltages and frequencies and scaling up the frequency does not often help reduce the energy consumption \cite{mei2016survey, bridges2016understanding}. 
\begin{equation}
P_{dynamic} = aCV^2f
\label{eq:dyn_power}
\end{equation}
Modern GPUs have two main frequency domains. One is core frequency which mainly controls the speed of stream multiprocessors (SMs) while the other is memory frequency which affects the bandwidth of DRAM. Table \ref{table:freq_parts} summarizes the dominating frequency for different types of memory. Note that only DRAM works under memory frequency and L2 cache works under core frequency though they both serve the global memory requests. 
\begin{table}[ht]
	\centering
	\caption{Dominating Frequency for different components.}
	\label{table:freq_parts}
	\begin{tabular}{lllll}
		\toprule
		Components  &  Dominating Frequency \\\midrule
		DRAM   &   memory frequency \\
		L2 Cache        &   core frequency     \\
		Shared Memory   &   core frequency     \\
		Texture Cache        &   core frequency     \\
		Register        &   core frequency     \\
		\bottomrule
	\end{tabular}
\end{table}
\subsection{Performance Scaling Behaviors with Frequency}
As different GPU applications may have various utilizations of different hardware components, changing the frequencies may lead to diverse performance scaling behaviors for them. As a motivating example, we test a set of frequency pairs on six GPU kernels to observe how the execution time changes. 

\begin{figure}[htbp]
	\centering     
	\subfigure[]
	{
		\includegraphics[width=0.46\linewidth]{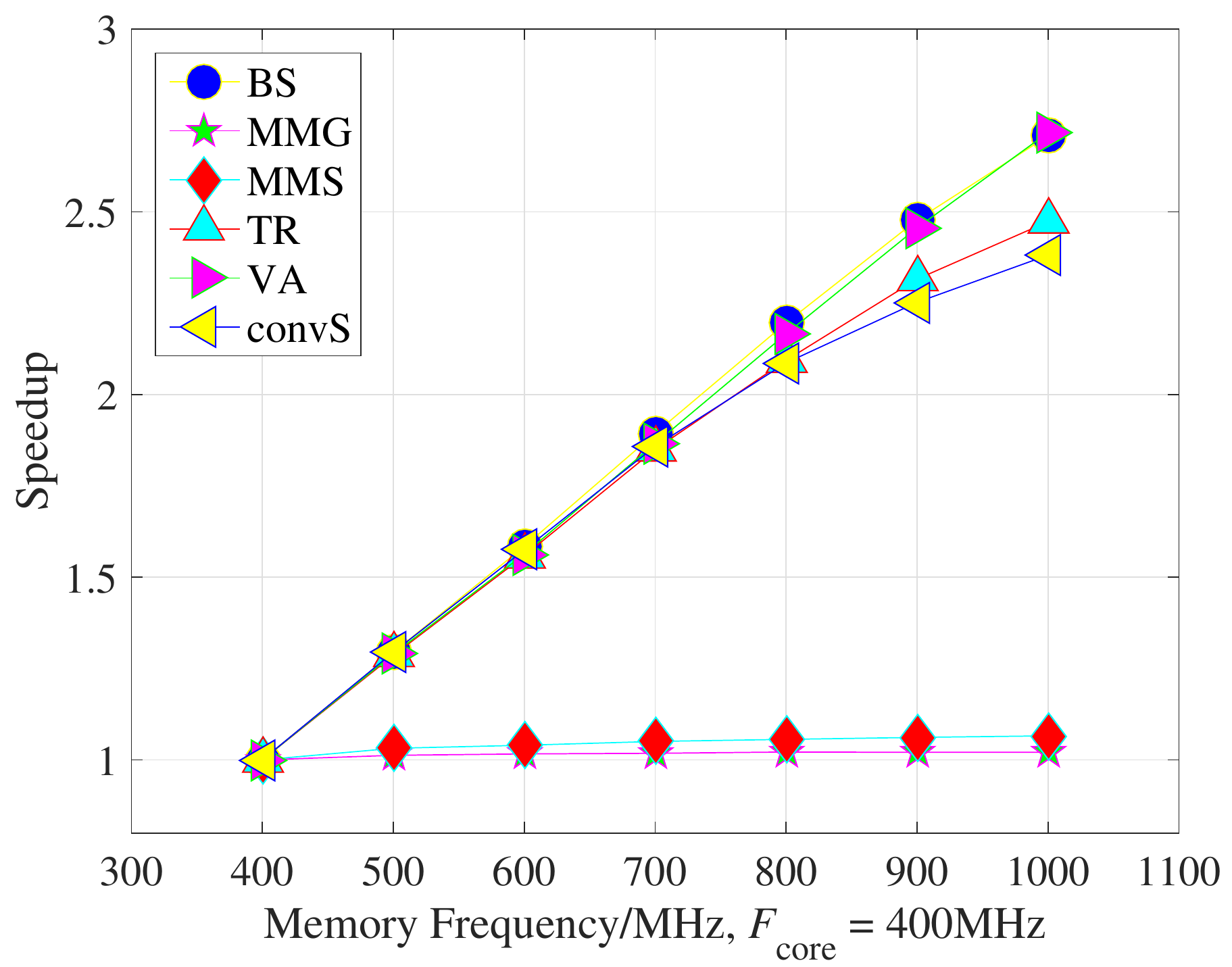}
		\label{fig:psa}
	}
	\subfigure[]
	{
		\includegraphics[width=0.46\linewidth]{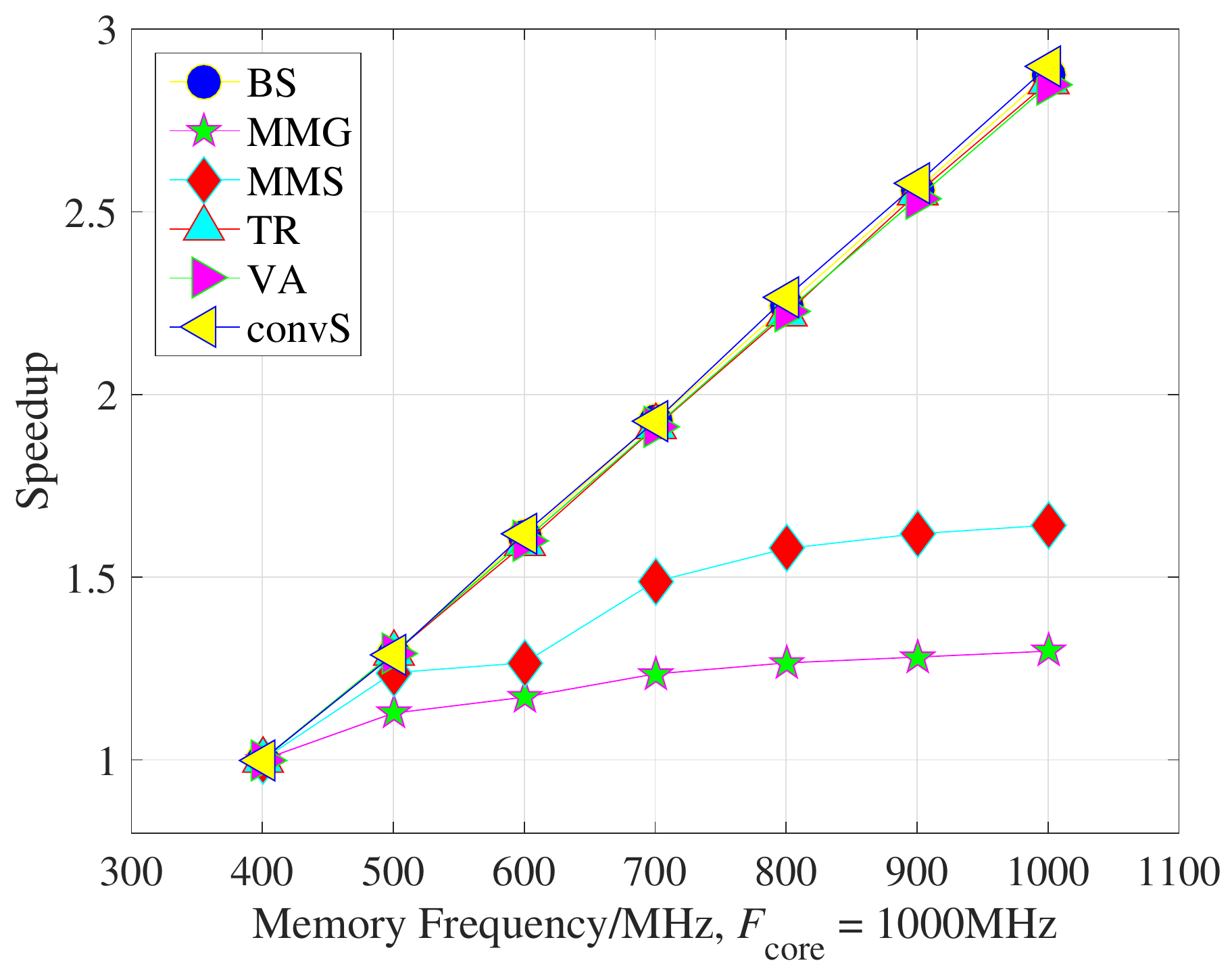}
		\label{fig:psb}
	}
	\subfigure[]
	{
		\includegraphics[width=0.46\linewidth]{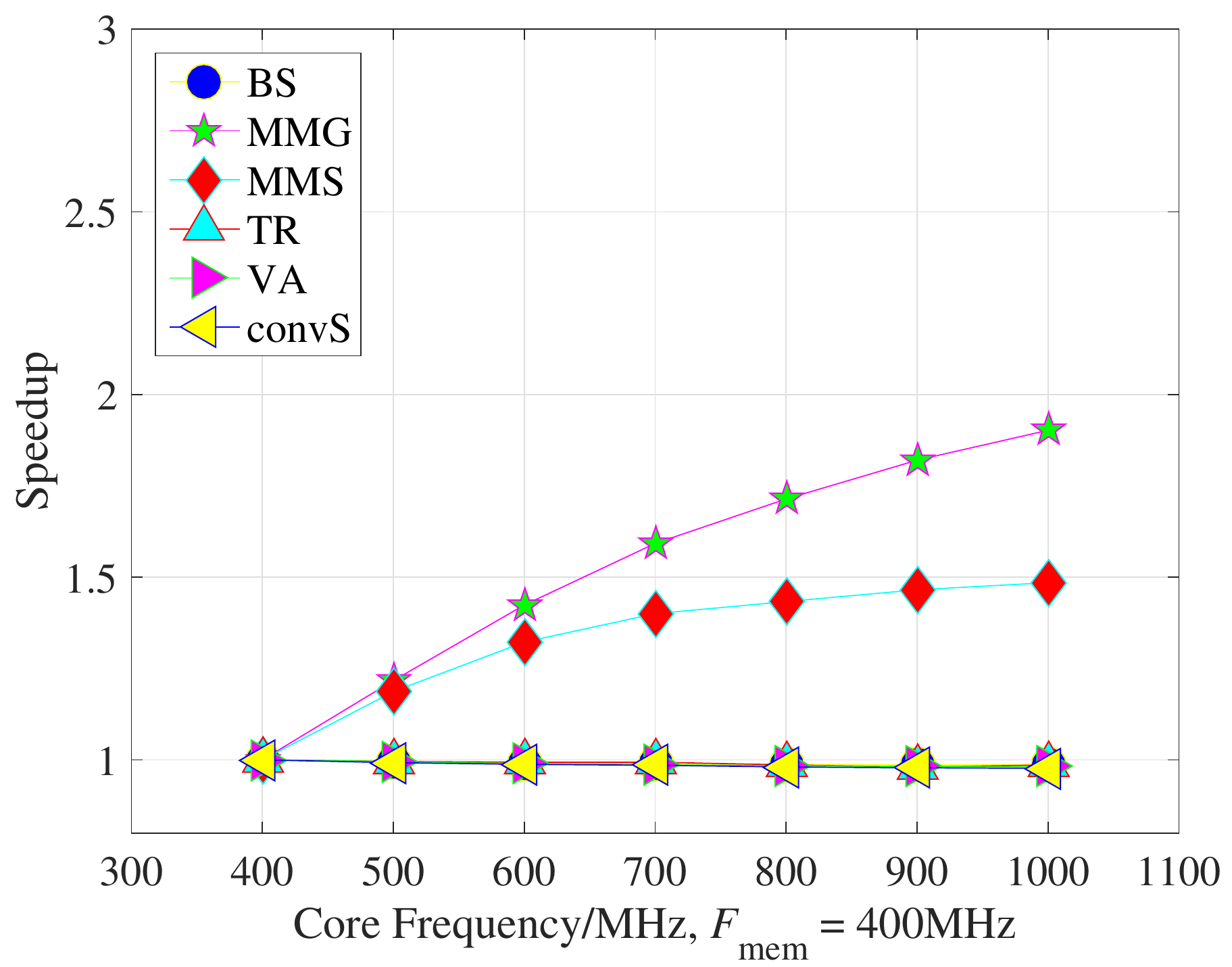}
		\label{fig:psc}
	}
	\subfigure[]
	{
		\includegraphics[width=0.46\linewidth]{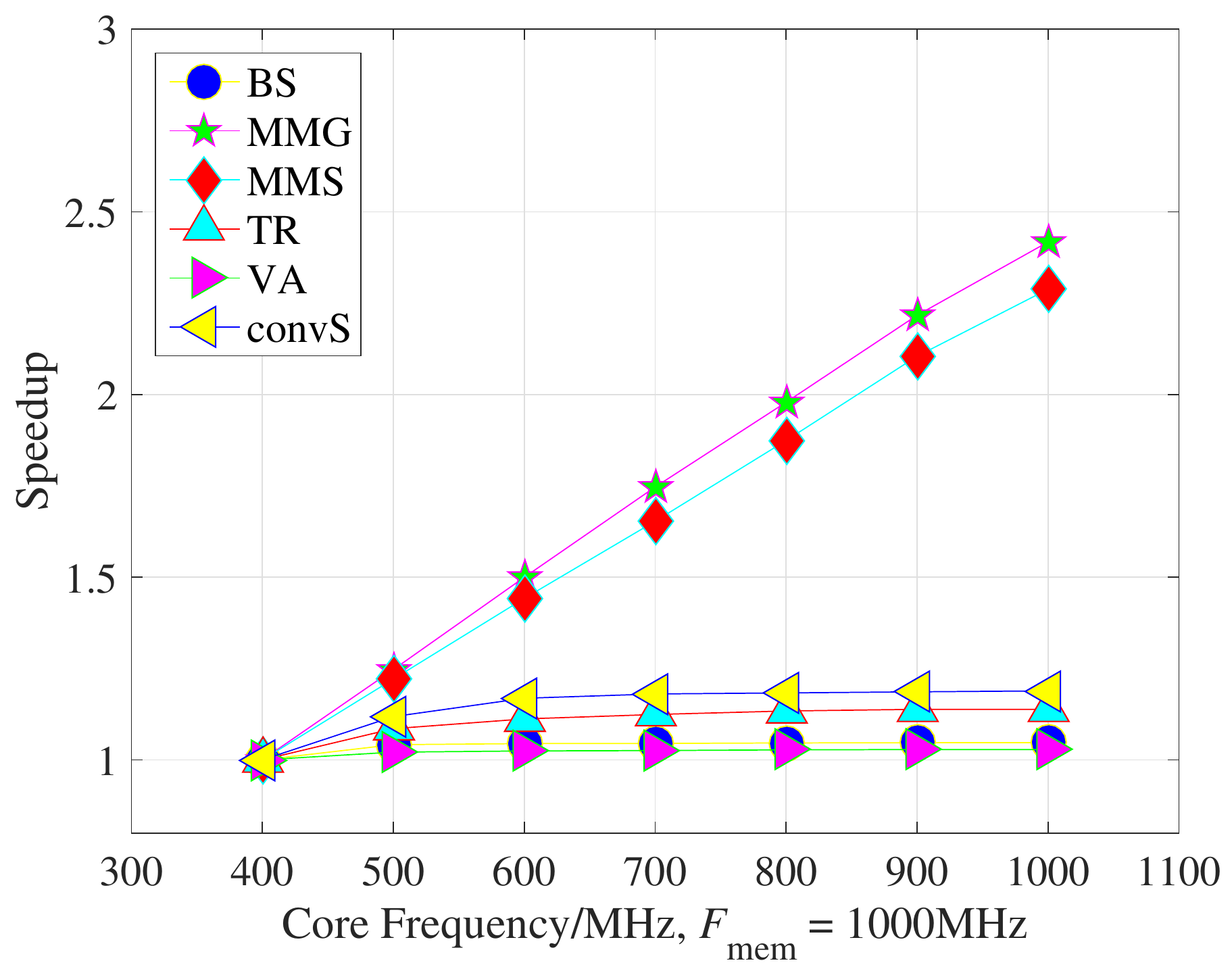}
		\label{fig:psd}
	}
	\caption{Performance scaling behavior under different frequency settings. The upper two figures show the speedup of different GPU kernels when increasing memory frequency with fixed core frequency. The below two figures show the speedup of different GPU kernels when increasing core frequency with fixed memory frequency.}
	\label{fig:perfScaling}
\end{figure}

We first fix the core frequency to 400 MHz and 1000 MHz respectively and scale the memory frequency from 400 MHz to 1000 MHz with a step size of 100 MHz. Illustrated by Fig. \ref{fig:psa} and \ref{fig:psb}, some kernels like transpose (TR), blackScholes (BS), vectorAdd (VA) and convolutionSeparable (convS) have almost over 2.5x speedup by increasing 2.5x memory frequency, while the other two matrix multiplication with global memory (MMG) and with shared memory (MMS) have negligible speedup. Another intesting finding is that two matrix multiplication kernels MMG and MMS have different scaling behaviors under different core frequency. Higher core frequency allows them to have higher speedup when increasing the memory frequency. The possible reason is that the performance is restricted by core frequency when core frequency is low while the performance is restricted by memory frequency when core frequency is high enough to drive the computational power.

Then we fix the memory frequency to 400 MHz and 1000 MHz respectively, and scale the core frequency from 400 MHz to 1000 MHz. Fig. \ref{fig:psc} and \ref{fig:psd} show that core frequency has little effects on the performance of TR, BS and VA but great impacts on the other three. It is also observed that the performance can be limited by different frequency domain with different frequency settings. 

As a result, under different frequency settings, the performance scaling behaviors can be diverse and complicated among different GPU kernels. Our goal is to establish an estimation model that can predict the execution time of a given GPU kernel under different core and memory frequency settings. In order to achieve it, we first explore how core and memory frequencies affect different types of memory including DRAM, L2 cache and shared memory. That gives us a quantitative model to estimate different memory latencies. Then we use profiling tools to extract some performance counters from running the kernel under the baseline frequency settings. Collaborated with the memory model and the profiling data, the kernel execution time can be estimated under other frequency settings.

\section{Related Work} \label{sec:RW}
To derive an accurate performance model of GPUs, it is quite important to understanding its complex memory hierarchy. Henry Wong et al. \cite{wong2010demystifying} and develop a micro-benchmark suite and measure some characteristics such as cache structure and latency of various memory types, TLB parameter, latency and throughput of arithmetic and logic operations. Meltzer \cite{meltzer2013micro} extended similar work on Tesla C2070. In addition, Xinxin Mei et al. \cite{mei2014npc, mei2015tpds} also conduct similar dissection but address more on memory hierarchy. They propose a fine-grained P-chase method to explore the cache parameters with uncommon structure and replacement policy which appears in the latest generation of GPU (Kepler and Maxwell). However, such methods usually test a single kernel with only few threads executing one type of instructions. When thousands of threads access memory simultaneously, which generally happens in GPU applications, the memory bandwidth might not satisfy the demands and some operations would be stalled in the queue of memory controller (MC). Such cases lead to high variance in memory access latency.

Hong et al. \cite{hong2009analytical, hong2010integrated} proposed an analytical model by estimating different degree of memory parallelism and computation parallelism with some offline information of the kernel program. Furthermore, Sim et al. \cite{sim2012performance} improves the above MWP-CWP model by considering cache effects, SFU characteristics and instruction throughput. However, their methods ignore the effects of shared memory latency and DRAM memory latency divergence, which may bring some significant biases in some memory-bounded application. Song et al. \cite{song2013simplified} extend their models and address more on different types of memory access by collecting some simple counters. However, the model averages the cache effects among all the warps and potentially ignores memory latency divergence in some asymmetric applications. 

Nath et al. \cite{nath2015crisp} present CRISP model which analyze the performance in the face of different frequencies of compute cores. They pointed out that DVFS on GPU is different from that on CPU since computation operations and memory transactions from different threads can overlap in most of the time. Based on the characteristics of GPU performance with varying frequencies found from experiments, they classify different execution stages in the kernel program and compute them with various frequencies. However, CRISP only works for the case of either scaling down the core frequency or scaling up the memory frequency. Also the model may be more complicated if considering the memory frequency. Joao et al. \cite{hpca2018power} proposed a GPU power estimation model with both core and memory frequency scaling. They designed well crafted microbenchmarks to extract the model parameters of each GPU components under the default frequency setting. Then they attempted to predict the power consumption of an application under over a wide range of frequency scaling. 

Recent state-of-art works witness the advantages of machine learning methods on GPU performance and power modeling. Gene Wu et al. \cite{wu2015gpgpu} built a performance model based on different patterns of scaling with various core frequency and memory frequency. He firstly adopted K-means to cluster different patterns of scaling behavior among 37 kernels and then explored the relationship between performance counters and clustering patterns with ANN modeling. With the model trained with large amount of data, one can predict the performance of one kernel under any setting of core frequency and memory frequency with the predicted scaling pattern. Wang et al. \cite{greenmetrics2017power} adopted SVR algorithm to estimate GPU power consumption considering both core and memory frequency scaling.

\section{Memory Modeling with Frequency Scaling} \label{sec:mem_model}
In the previous performance modeling work, memory latency is usually set as a constant parameter obtained by micro-benchmarking. However, since the DRAM in the GPUs can be accessed by any threads running on any SMs, the memory latency of each thread may vary due to intensive memory transactions. Besides, memory latency can also change with different frequency settings. In this section, we first model the memory latency with a simple queueing model. Then we measure the parameters used in the model with different frequency settings for further performance modeling. 
\subsection{DRAM latency} 
\begin{figure}[ht]
	\centering
	\advance\leftskip0.01\textwidth
	\includegraphics[width=0.47\textwidth]{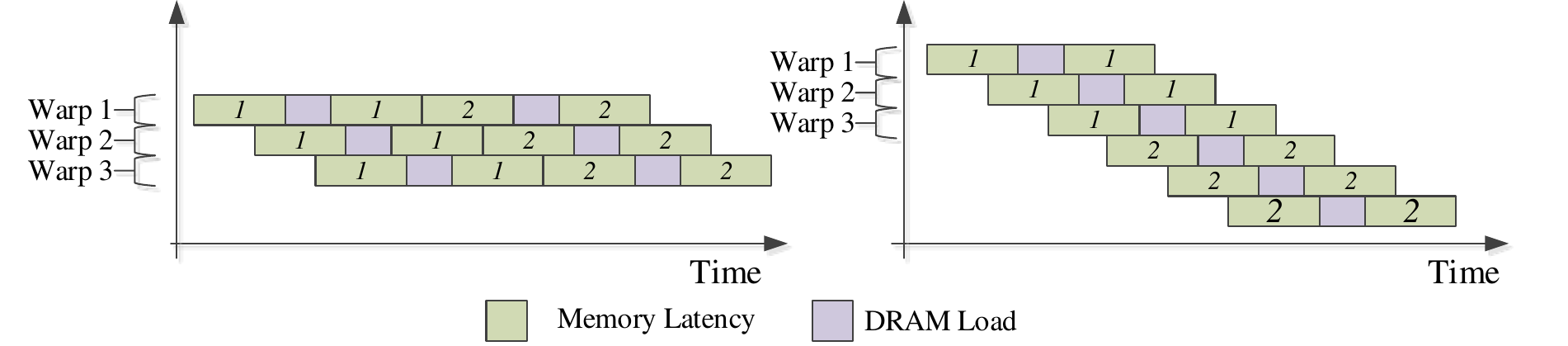}\\
	\caption{Execution time pipeline of infrequent DRAM requests. The number on the block indicates the iterations of one warp.}
	\label{fig:mem_not_full}
\end{figure}

\begin{figure}[ht]
	\centering
	\advance\leftskip0.01\textwidth
	\includegraphics[width=0.47\textwidth]{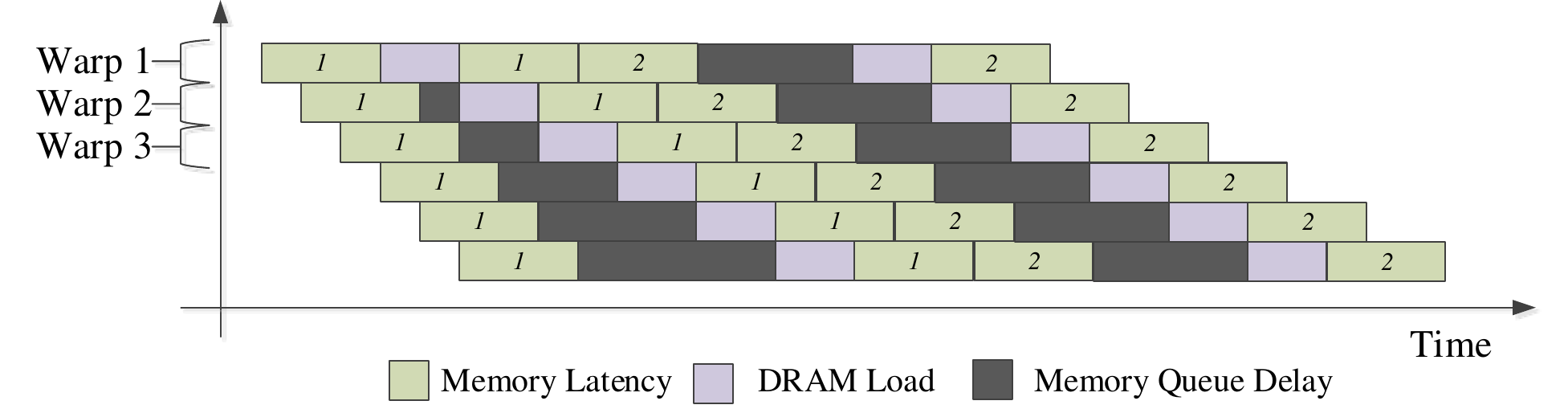}\\
	\caption{Execution time pipeline of intensive DRAM requests. The number on the block indicates the iterations of one warp.}
	\label{fig:mem_saturated}
\end{figure}

When one warp launches a memory request to the global memory, it usually takes about hundreds of cycles to go through DRAM if the data is not cached. This minimum latency happens when the memory system is idle and only contains the overhead of path traveling and data access. Fig. \ref{fig:mem_not_full} shows this case. The inter-arrival intervals between two consequent memory requests is shorter than the time consumption of loading data from DRAM. Thus, each memory request only costs the minimum latency. To compute the total latency of finishing all the memory requests in this case, we only need to care how many memory requests are executed by a warp. Inferred from Fig. \ref{fig:mem_not_full}, the total time consumption $T\_lat$ of all the memory requests can be calculated by Equation \ref{eq:total_lat_not_full}. $interArr$ denotes the inter-arrival time of two consequent memory requests. $\#W$ denotes the total warp number. $dm\_lat$ denotes the minimum memory latency with no memory contention. $gld\_trans$ denotes the global memory transactions of one warp. 
\begin{equation}
T\_lat = interArr \times \#W + dm\_lat \times gld\_trans
\label{eq:total_lat_not_full}
\end{equation}
If the memory system is saturated due to the intensive memory requests, the minimum latency can hardly be achieved. Most requests should waiting in the queue until the previous ones have been finished. In Fig. \ref{fig:mem_saturated}, each memory request is launched with a very short interval so that each memory request not only takes the minimum latency but also the queueing delay, which means the waiting time in the queue. Thus, intensive memory access demands can lead to diverse memory latency. In this case, we can calculate the total time consumption by Equation \ref{eq:total_lat_sat}. $dm\_del$ denotes the service time of one memory transaction. 
\begin{equation}
T\_lat = dm\_lat + dm\_del \times gld\_trans \times \#W
\label{eq:total_lat_sat}
\end{equation}
\begin{figure}[htbp]
	\centering     
	\subfigure[]
	{
		\includegraphics[width=0.22\textwidth]{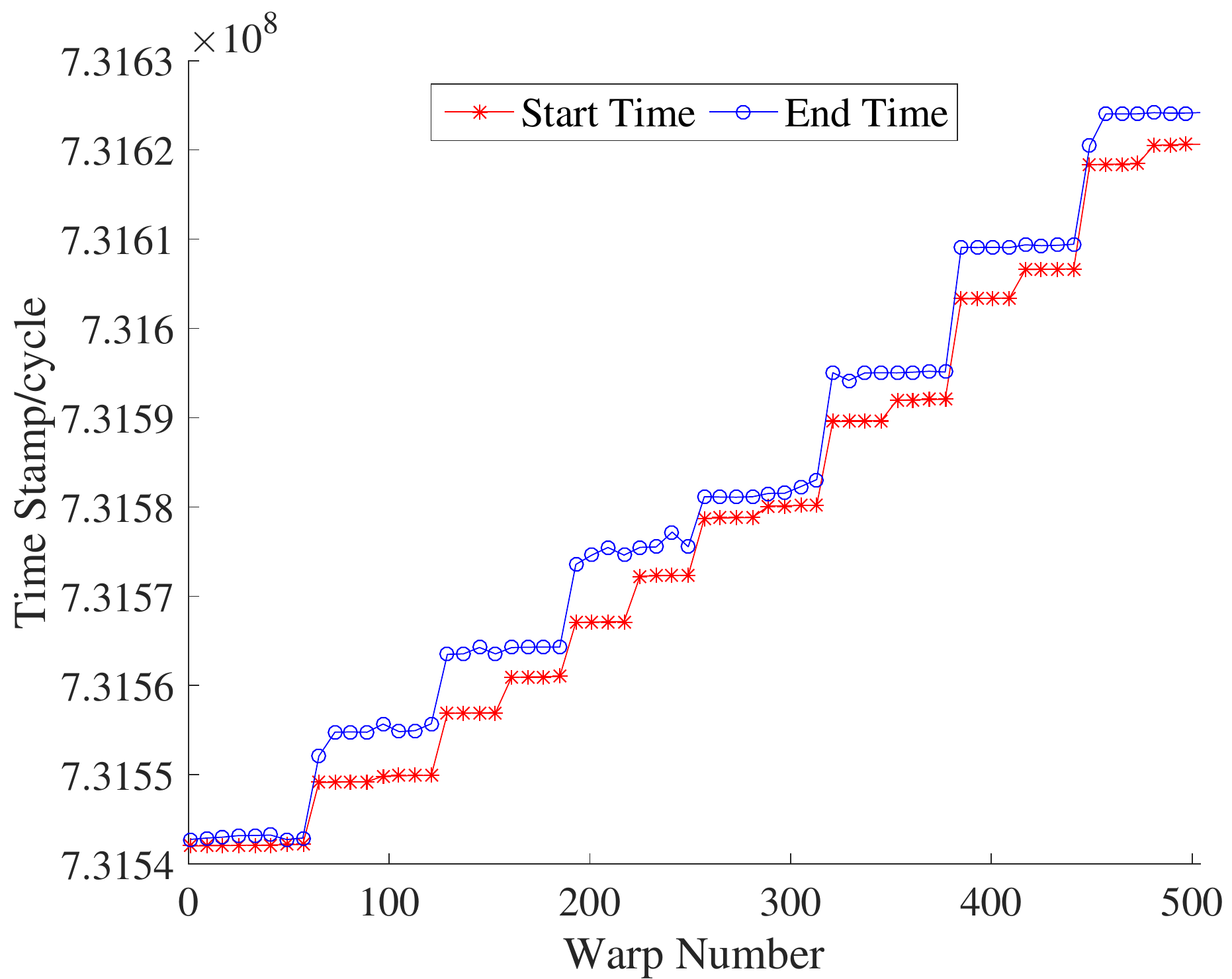}
		\label{fig:mb_wq}
	}
	\subfigure[]
	{
		\includegraphics[width=0.22\textwidth]{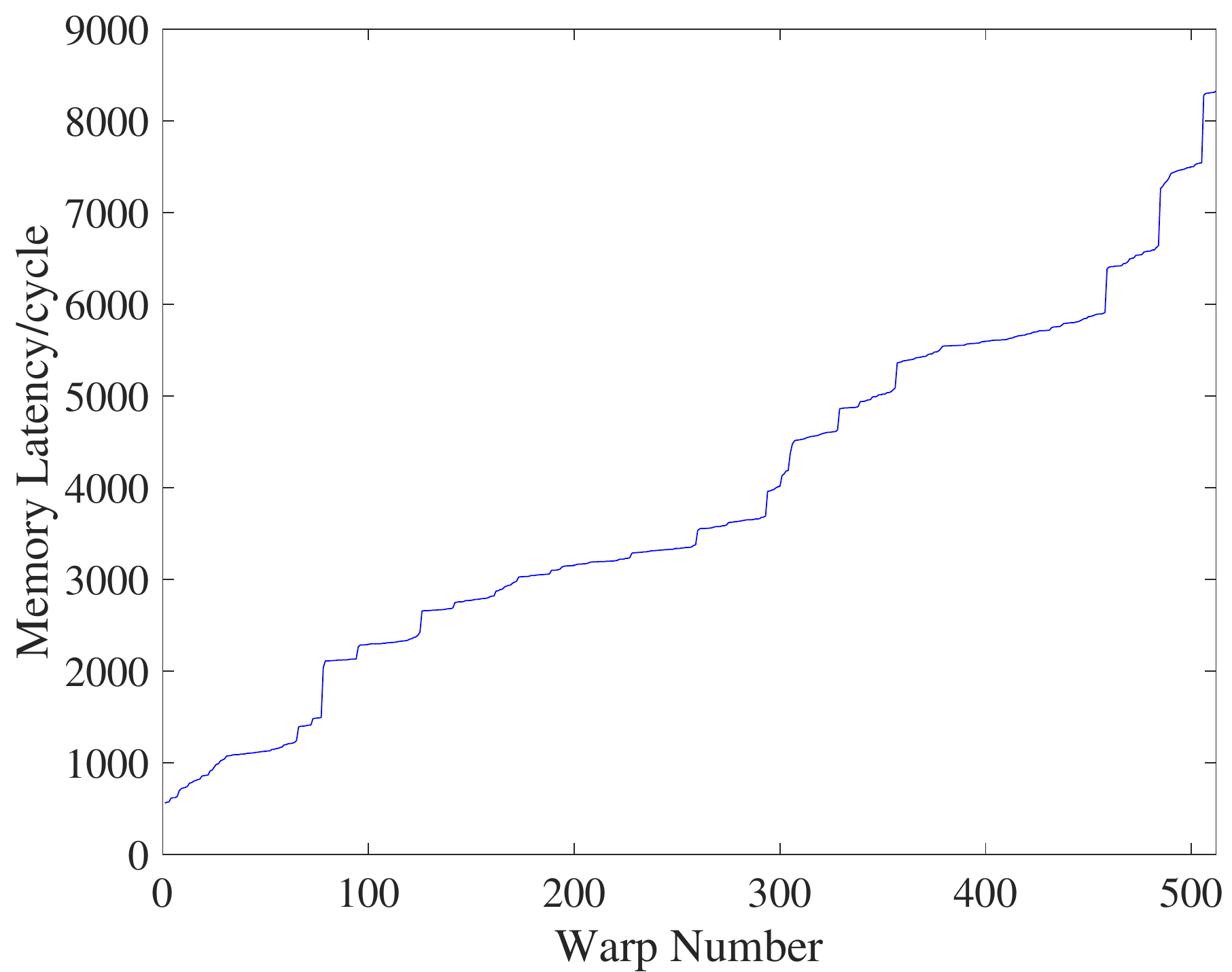}
		\label{fig:mb_mlo}
	}
	\caption{Experimental results of memory access latency. The time stamp data in \ref{fig:mb_wq} is sorted by starting time in ascending order. The memory access latency of each warp in \ref{fig:mb_mlo} is ascendingly re-ordered.}
	\label{fig:mem_benchmark}
\end{figure}
We also observe this phenomenon in experiments. We revise the global bandwidth benchmark code of \cite{mei2015tpds} and add clock measuring function $clock()$ to collect memory latency samples. To reduce as much overhead of $clock()$ and time recording in global memory as possible, we sample only one request for some threads. Fig. \ref{fig:mem_benchmark} shows our experimental results. We can infer two things from the results. First, the memory latency can be diverse because of the intensive requests. Second, the memory latency is somehow linearly correlated with warp numbers, which meets the model in Fig. \ref{fig:mem_saturated}. 
%
%
Then we would like to explore how frequency scaling affects $dm\_lat$ and $dm\_del$. We first utilize the global latency benchmark code of \cite{mei2015tpds} to measure the $dm\_lat$ under different memory frequencies. For simplicity, we only measure the latency in the case that the TLB cache is hit. Table \ref{tab:dram_lat_mem_scaling} shows part of our results. The cycle is the time unit under GPU core frequency. We also measure $dm\_lat$ with other frequency combinations and find that it can be fitted by Equation \ref{eq:min_lat} with 0.9959 R-squared.
\begin{equation}
dm\_lat = 222.78 \times core\_f/mem\_f + 277.32
\label{eq:min_lat}
\end{equation}
As for DRAM delay measurement, we also use the global bandwidth benchmark code of \cite{mei2015tpds} to achieve as high DRAM bandwidth as possible. Table \ref{tab:dram_del_mem_scaling} shows part of our results. To calculate $dm\_del$ in cycles, we transfer the total execution time into cycles by multiplying the time and the core frequency and calculate how many memory transactions of all warps. Then we can infer $dm\_del$ by Equation \ref{eq:total_lat_sat} with the obtained $dm\_lat$. We observe the DRAM delay is also correlated with the bandwidth efficiency, which means the percentage of DRAM bandwidth utilization. By increasing memory frequency, the cycles of $dm\_del$ become smaller and the bandwidth efficiency becomes larger, which suggests that high memory frequency helps improve utilization of DRAM bandwidth. 
\begin{table}[ht]
	\centering
	\caption{Minimum DRAM latency under different memory frequencies}
	\label{tab:dram_lat_mem_scaling}
	\scriptsize{
		\begin{tabular}{|p{0.5in}|p{0.5in}|p{0.4in}|} \hline
			\textbf{Memory Freq./MHz} & \textbf{Core Freq./MHz} & Cycles		\\ \hline
			400& 400 & 500		\\ \hline		
			500 & 500 & 455.5 		\\ \hline	
			600 & 600 & 425.8  	\\ \hline		
			700 & 700 & 404.6  \\ \hline	
			800 & 800 & 388.7  	\\ \hline
			900  & 900  &  376.3   \\ \hline		
			1000  & 1000  &  366.4   \\ \hline		
		\end{tabular}
	}
\end{table}
\begin{table}[ht]
	\centering
	\caption{DRAM read delay under different memory frequencies}
	\label{tab:dram_del_mem_scaling}
	\scriptsize{
		\begin{tabular}{|p{0.5in}|p{0.5in}|p{0.3in}|p{0.5in}|} \hline
			\textbf{Memory Freq./MHz} & \textbf{Core Freq./MHz} & Cycles & Bandwidth Efficiency		\\ \hline
			400& 400 & 10.06 & 76\%	\\ \hline		
			500 & 500 & 9.76 & 78.13\%		\\ \hline	
			600 & 600 & 9.54 & 79.8\% 	\\ \hline		
			700 & 700 & 9.31 & 81.83\%	\\ \hline	
			800 & 800 & 9.19 & 83.42\% 	\\ \hline
			900  & 900  & 9.06 & 84.51\%  \\ \hline		
			1000  & 1000  & 9.0 & 85\%  \\ \hline		
		\end{tabular}
	}
\end{table}
\subsection{L2 cache}
With the development of GPU generations, L2 cache becomes larger and larger (e.g. from 512 KB for Fermi GTX560Ti to 2 MB for Maxwell GTX980) in order to reduce the pressure of memory system. As mentioned before, different cache hit rate may bring different sensitivity to frequency scaling. Similar with the DRAM measurement experiments, we also use the same global latency code to obtain L2 cache latency under different frequency settings. We observe that the latency is always within 220 to 224 cycles. This is reasonable because L2 cache is affected by core frequency. Thus, we take the average 222 cycles for L2 minimum latency. Besides, we choose 1 cycle for $l2\_del$ due to the truth that L2 cache can return a memory request per core cycle with the same reason. 

\subsection{Adjustment with frequency scaling}
For simplicity, our performance model defines baseline frequency setting in which the ratio of memory frequency to core frequency is one. We measure some basic latency and throughput for all the components including core computation, shared memory access, constant memory, L2 cache and DRAM under baseline setting. We can use the standard Average Memory Access Time (AMAT) \cite{hennessy2011computer} model to obtain average global memory access latency $agl\_lat$ and average queueing delay $agl\_del$ of all the global memory transactions happenin during kernel execution with Equation \ref{eq:agl_lat_a} and \ref{eq:agl_lat_b}. $l2\_hr$ denotes L2 cache hit rate of the kernel. $core\_f$ and $mem\_f$ denotes the frequencies of core and memory respectively. 
\begin{subequations}
	\label{eq:agl_lat}
	\begin{align}
	agl\_lat = &l2\_lat \times l2\_hr + dm\_lat \nonumber	\\
	& \times (core\_f / mem\_f) \times (1 - l2\_hr)	\label{eq:agl_lat_a}\\
	agl\_del = &l2\_del \times l2\_hr + dm\_del \nonumber	\\
	& \times (core\_f / mem\_f) \times (1 - l2\_hr)    \label{eq:agl_lat_b} 
	\end{align}
\end{subequations}

Since we calculate the execution time in the scope of core frequency, there is no extra adjustment to those latency and throughput in SM. 

%

\section{GPU Performance Modeling with Frequency Scaling} \label{sec:GPM}
The SMs in GPUs execute threads in groups of 32 parallel threads called warps \cite{cudaProgram}. Generally GPU can launch a large amount of warps during the whole kernel execution period. However, due to the hardware resource limitation, one SM can only executes a certain number (denoted by \#Aw) of warps concurrently called active warps, denoted by \#Aw. Once we obtain the time consumption of a round of active warps, denoted by $T_{active}$, the total execution time of a kernel can be estimated by Equation \ref{eq:time_exec}. \#B denotes the total number of thread blocks; \#Wpb denotes the number of warps per block; \#SM denotes the number of SMs.
\begin{equation}
T_{exec} = T_{active} * (\#Wpb * \#B / (\#Aw \times \#SM))
\label{eq:time_exec}
\end{equation}
Basically, a GPU kernel can be divided into several segments. Some segment does not not access shared memory, while some does. Since they are influenced by different frequencies and usually have different working patterns, we classify the GPU kernels segments into two categories by whether shared memory transactions happen during the kernel execution. Some kernels also utilize texture/L1 cache for further performance improvement and somehow affect the accuracy of our model. We will leave it as future improvement work for our current model. 
\subsection{Performance Modeling without Shared Memory}
\begin{figure} [ht]
	\advance\leftskip0.01\textwidth
	\includegraphics[width=0.45\textwidth]{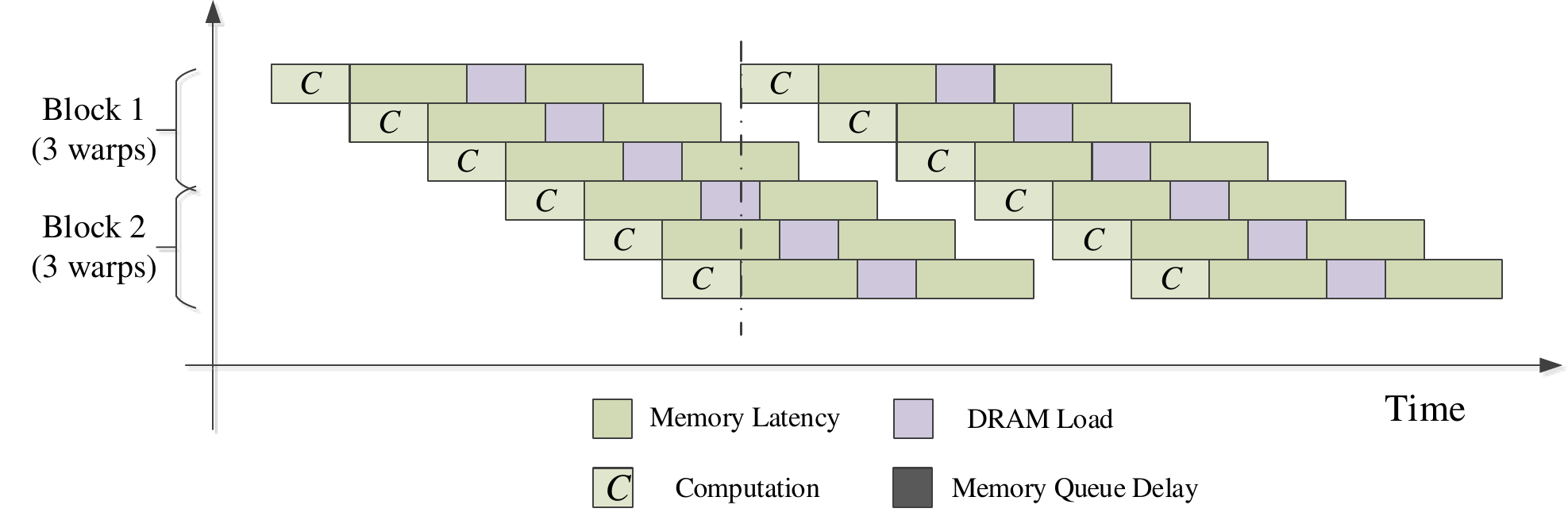}\\
	\caption{Execution time pipeline of a compute-dominated kernel. Since the kernel launches enough warps containing long compute cycles, most of the memory latency can be hidden.}
	\label{fig:pipeline_comp}
\end{figure}

The first case of our model is that the GPU kernel does not utilize shared memory. In this case, the kernel only contains computation in SMs and global memory transactions. As described in Section \ref{sec:mem_model}, we can estimate the time consumption of global memory transactions. As for computation part, we simply assume that there happens the same computation time before each global memory transaction as shown in Fig. \ref{fig:pipeline_comp}. We divide the total compute instructions (denoted by $comp\_inst$) by total global memory transactions $gld\_trans$ to obtain average compute instructions number (denoted by $avr\_inst$). The average computation time (denoted by $avr\_comp$) before each global memory transaction can be estimated by Equation \eqref{eq:av_com_a} and \eqref{eq:av_com_b}.
\begin{subequations}
	\label{eq:av_com}
	\begin{align}
	avr\_inst = comp\_inst / gld\_trans  \label{eq:av_com_a}\\
	avr\_comp = inst\_cycle \times avr\_inst  \label{eq:av_com_b}
	\end{align}
\end{subequations}
If the kernel launches large number of computation instructions and only few memory requests which do not saturate the memory bandwidth, the computation period will dominate the total kernel execution time. On the other hand, if the memory access latency is somehow much longer than computation cycle due to intensive memory requests, the computation period can be hidden by memory operations. The first case is regarded as compute-dominated or compute-bound kernel while the second is memory-dominated or memory-bound. 
\begin{figure}[ht]
	\advance\leftskip0.01\textwidth
	\includegraphics[width=0.45\textwidth]{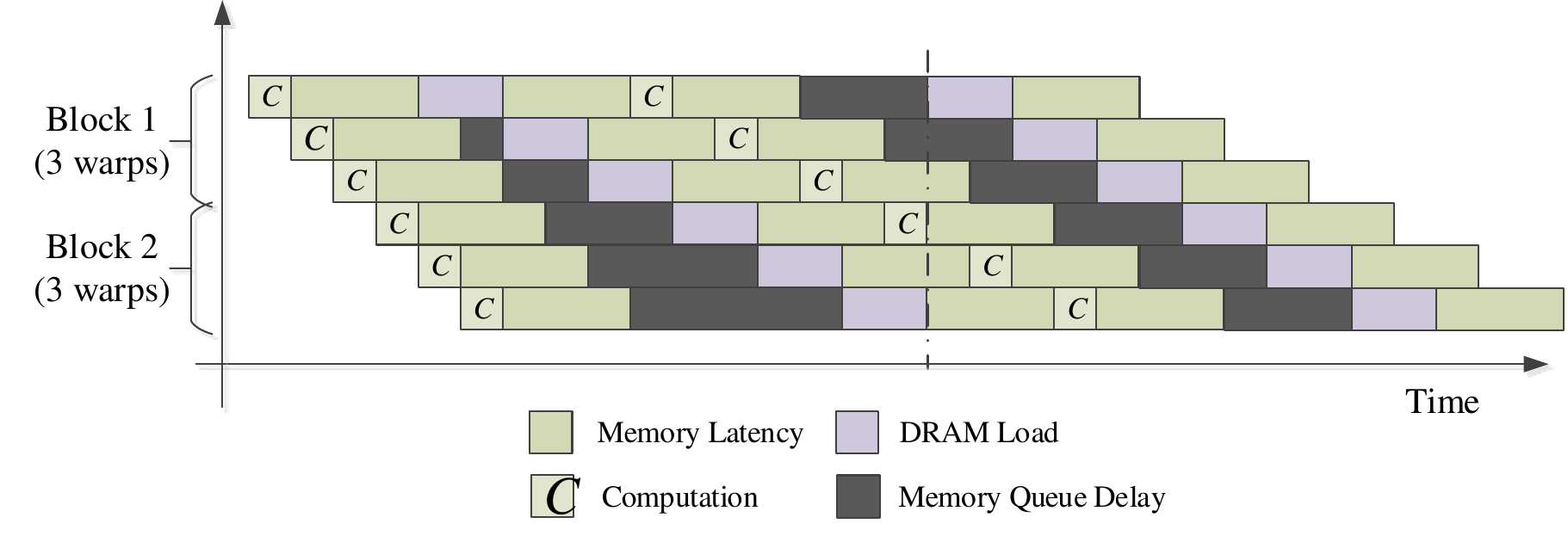}\\
	\caption{Execution time pipeline of a memory-intensive kernel. Since each warp issues a few computation instructions but frequent memory access requests, one memory transaction cannot be processed until all outstanding transactions have been finished.}
	\label{fig:pipeline_mem}
\end{figure}
\subsubsection{\textbf{Compute-Dominated}}
When there are enough computation instructions to be issued and the memory requests are not too intensive due to long computation period, the global memory latency can be hidden, as illustrated by Fig. \ref{fig:pipeline_comp}. In this case, Equation \eqref{eq:comp_cond1_a} and \eqref{eq:comp_cond1_b} should be satisfied. We can estimate $T_{active}$ by Equation \eqref{eq:comp-case-cycle}. $o\_itrs$ denotes the repeat times of one computation period and one global memory transaction. 

\begin{subequations} \small
	\label{eq:comp_cond1}
	\begin{align}
	&avr\_comp \geq agl\_del \label{eq:comp_cond1_a}\\
	(&avr\_comp \times (\#Aw - 1)) \geq agl\_lat  \label{eq:comp_cond1_b}
	\end{align}
\end{subequations}
\begin{align} \label{eq:comp-case-cycle}
	T_{active} = &avr\_comp \times \#Aw \times o\_itrs + agl\_lat	
\end{align}

\subsubsection{\textbf{Memory-Dominated}}
When the memory bandwidth is saturated or there are not enough warps to issue computation instructions, one memory request is waiting until all the outstanding requests have been finished. Fig. \ref{fig:pipeline_mem} demonstrates this case. The condition is described as Equation \eqref{eq:mem_cond2_a} and \eqref{eq:mem_cond2_b}. Similar with the case in Fig. \ref{fig:mem_saturated}, we can regard the compute cycles as inter-arrival time of two consequent memory requests. We can estimate $T_{active}$ by Equation \eqref{eq:mem-case-cycle} by focusing on the $agl\_del$ of each warp.

\begin{subequations} \small
	\label{eq:mem_cond2}
	\begin{align}
	&avr\_comp \leq agl\_del   \label{eq:mem_cond2_a}\\ 
						(&avr\_comp + agl\_lat) \geq (agl\_del \times (\#Aw - 1))  \label{eq:mem_cond2_b}
	\end{align}
\end{subequations}
\begin{align} \label{eq:mem-case-cycle}
T_{active} = &agl\_lat + avr\_comp \nonumber \\
&+ agl\_del \times \#Wpb \times o\_itrs
\end{align}

\begin{figure} [ht]
	\advance\leftskip0.01\textwidth
	\includegraphics[width=0.45\textwidth]{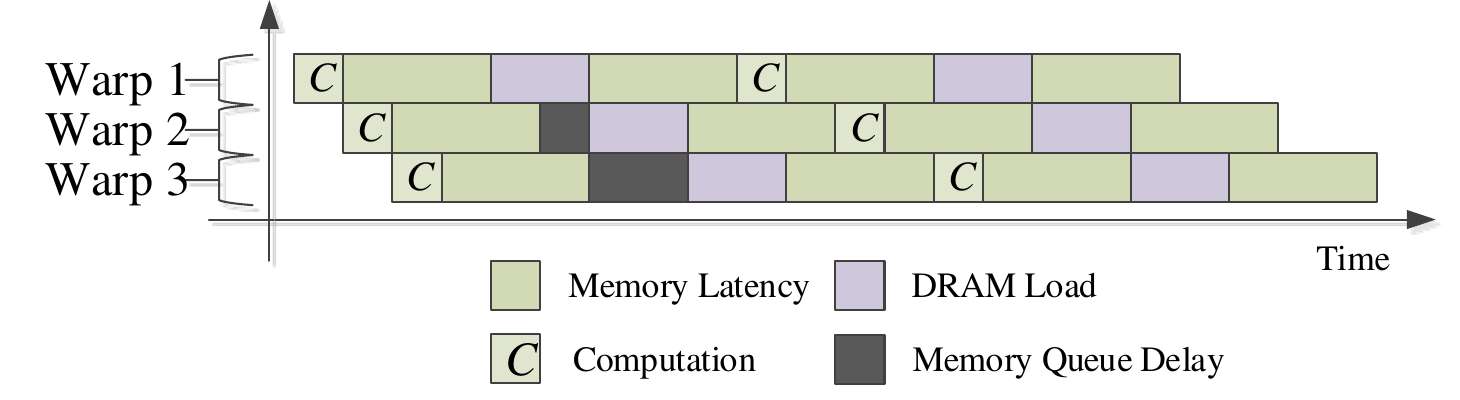}\\
	\caption{Execution time pipeline of a kernel containing few warps that have short computation periods. Thus, the first memory request of each warp has waiting period while the rest do not.}
	\label{fig:new-1}
\end{figure}

When the kernel launches only a few warps, most latency cannot be hidden which leads to insufficient utilizations of the GPU. The memory latency may contribute a lot to the execution time. There are two cases identified by whether $avr\_comp$ is shorter than $agl\_del$. When $avr\_comp$ is shorter than $agl\_del$, the first memory request of each warp should have waiting period as Fig. \ref{fig:new-1} shows. It can be described as Equation \eqref{eq:mem_cond3_a} and \eqref{eq:mem_cond3_b}. We can estimate $T_{active}$ by Equation \eqref{eq:new1-case-cycle} for this case.

\begin{figure} [ht]
	\advance\leftskip0.01\textwidth
	\includegraphics[width=0.45\textwidth]{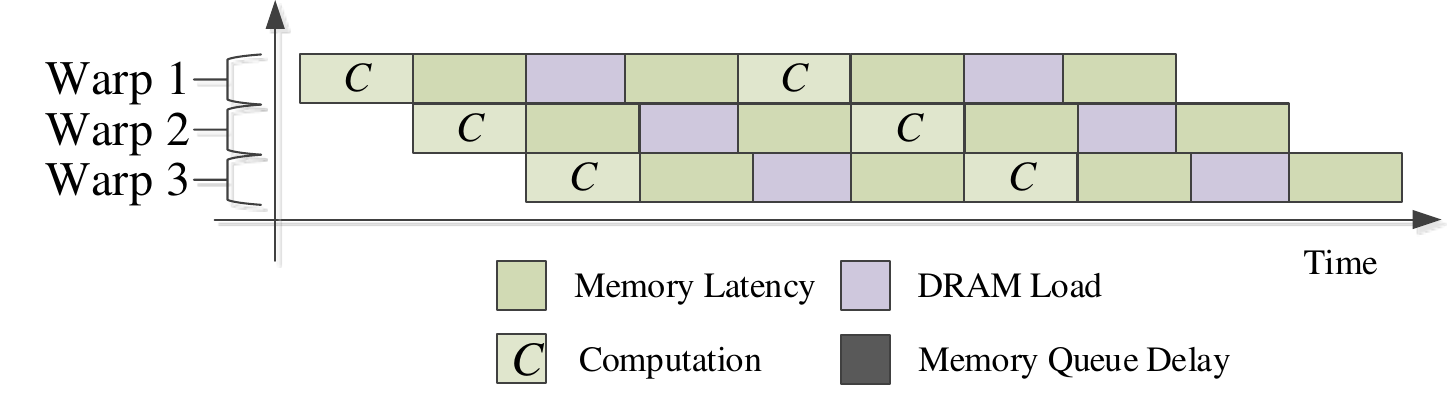}\\
	\caption{Execution time pipeline of a kernel containing few warps that have long computation periods. Thus, each memory request do not need to wait and can be processed immediately.}
	\label{fig:new-2}
\end{figure}
\begin{subequations} \small
	\label{eq:mem_cond3}
	\begin{align}
	&avr\_comp \leq agl\_del   \label{eq:mem_cond3_a}\\ 
	(&avr\_comp + agl\_lat) \leq (agl\_del \times (\#Aw - 1))  \label{eq:mem_cond3_b}
	\end{align}
\end{subequations}
\begin{align} \label{eq:new1-case-cycle}
T_{active} = &agl\_del \times \#Aw + agl\_lat + avr\_comp \nonumber \\
&+(avr\_comp + agl\_lat) \times (o\_itrs - 1) 
\end{align}
When $avr\_comp$ is longer than $agl\_del$, all the memory requests can be processed immediately once issued as Fig. \ref{fig:new-2} shows. It can be described as Equation \eqref{eq:mem_cond4_a} and \eqref{eq:mem_cond4_b}. We can estimate $T_{active}$ by Equation \eqref{eq:new2-case-cycle}.

\begin{subequations} \small
	\label{eq:mem_cond4}
	\begin{align}
	&avr\_comp \geq agl\_del   \label{eq:mem_cond4_a}\\ 
	(&avr\_comp \times (\#Aw - 1)) \leq agl\_lat  \label{eq:mem_cond4_b}
	\end{align}
\end{subequations}
\begin{align} \label{eq:new2-case-cycle}
T_{active} = &avr\_comp \times (\#Aw - 1) \nonumber \\
& + (avr\_comp + agl\_lat) \times o\_itrs 
\end{align}

\subsection{\textbf{Performance Modeling with Shared Memory}}
Shared memory plays an important role in GPU performance optimization since it has lower latency and higher throughput compared to DRAM and can be shared among the threads within the same block. Its latency and throughput is affected by core frequency. One GPU kernel may have different patterns of utilization of shared memory which makes the performance estimation complicated. Basically, there are three phases for this kind of kernels. At the beginning, all the warps send memory requests to global memory and then store the data into shared memory. Second, threads within the same block access shared memory for computation. Finally, the results are written back to global memory. According to the access workload to shared memory, we design the performance models for two cases as follows.
\begin{figure}[ht]
	\centering
	\includegraphics[width=0.45\textwidth]{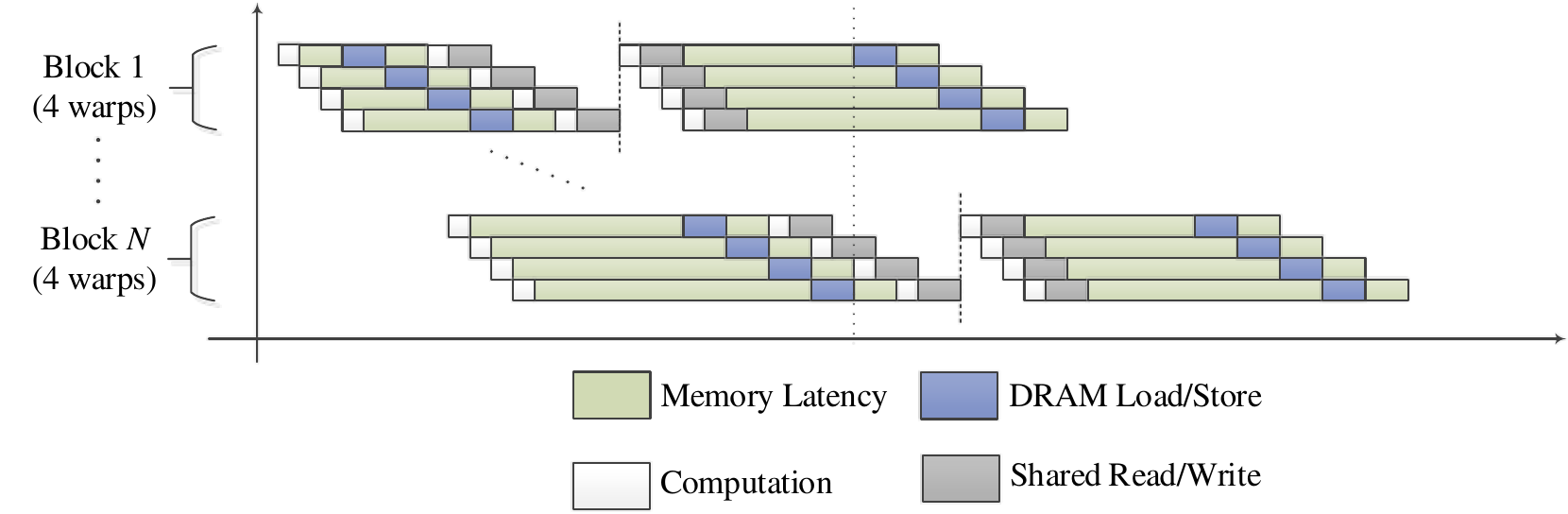}\\
	\caption{Execution time pipeline of a kernel containing infrequency shared memory access. Since there are very few shared memory transactions, the first block finishes them quickly and begins to send global memory request to DRAM while the final block even have not finished the first global memory transaction. }
	\label{fig:transpose_Coale}
\end{figure}
\subsubsection{\textbf{Shared memory requests are infrequent}}
Some kernels may only have few iterations of shared memory requests. In this case, since the latency of shared memory access is much lower than that of global memory access, the shared memory latency can often be hidden by global memory latency. Fig. \ref{fig:transpose_Coale} shows this case. In the first phase, all the warps are launching global memory requests and storing the data into shared memory, which results in heavy traffic in DRAM. In the second phase, each warp only executes one shared memory access consuming only a small number of cycles. Then each warp writes the results back to global memory, which again launchs quite a number of global memory transactions. The pattern is similar with Fig. \ref{fig:pipeline_mem} except that it is shared memory latency to be hidden. The condition is given by Equation \eqref{eq:shm_cond1_a} and \eqref{eq:shm_cond1_b} for this case and we can estimate the execution time by Equation \eqref{eq:shm-cycle-if}. $shm\_lat$ denotes shared memory latency. Transpose with coalesced optimization is one instance of this case. Since the shared memory latency can be hidden, the kernel is not sensitive to core frequency but memory frequency, which also meets the results of our previous motivating examples. 

\begin{subequations} \small
	\label{eq:shm_cond1}
	\begin{align}
	&avr\_comp \leq agl\_del   \label{eq:shm_cond1_a}\\ 
	(&avr\_comp + shm\_lat) \leq (agl\_del \times (\#Aw - \#Wpb))  \label{eq:shm_cond1_b}
	\end{align}
\end{subequations}
\begin{align} \label{eq:shm-cycle-if}
T_{active} = &avr\_comp + agl\_lat    \nonumber \\
& + agl\_del \times \#Aw \times gld\_trans  
\end{align}
\begin{figure}[ht]
	\centering
	\includegraphics[width=0.5\textwidth]{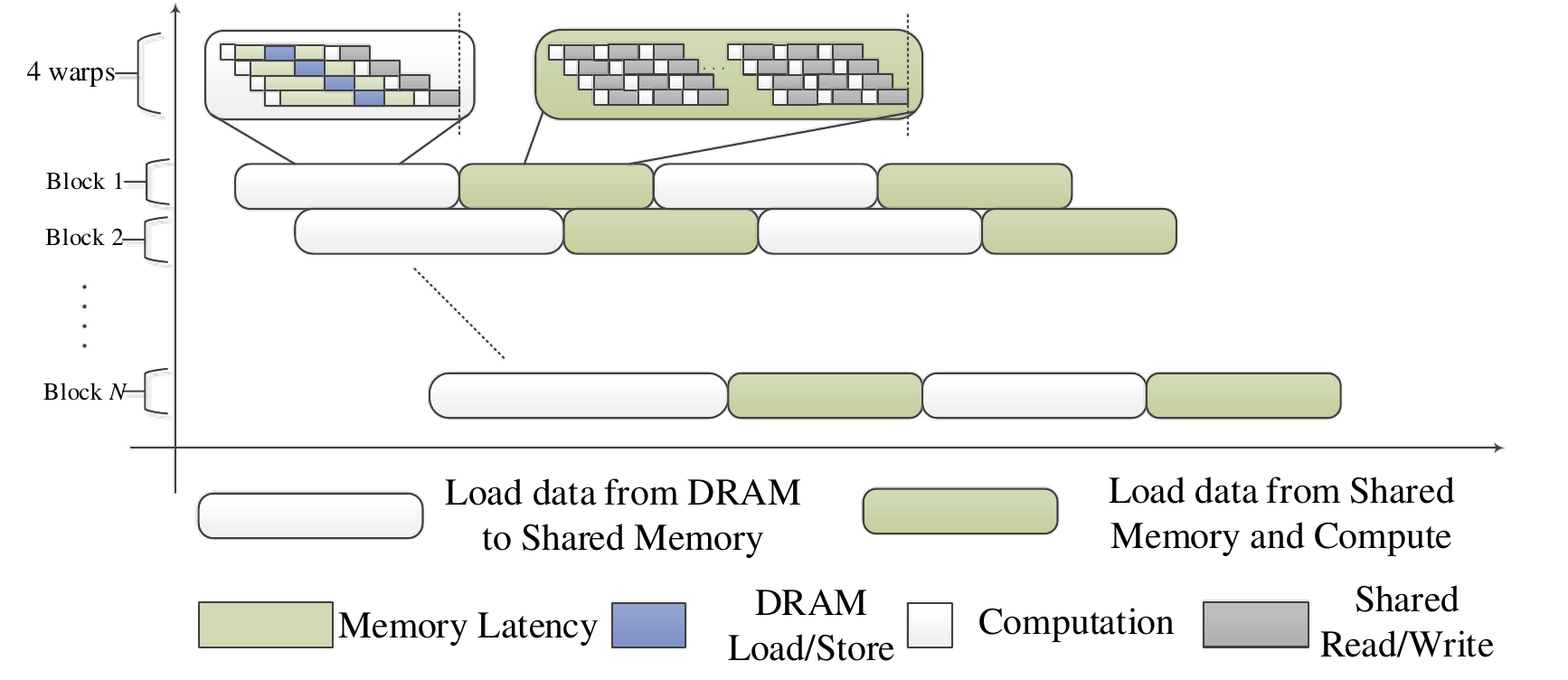}\\
	\caption{Execution time pipeline of matrix multiplication with shared memory. Phase 1 contains a large number of global memory requests. Phase 2 contains multiple shared memory operations. Since Phase 2 is long enough to hide the global memory latency of other blocks, the rest Phase 1 only has memory contention within the same block due to synchronization function.}
	\label{fig:matrixMul_ShMem}
\end{figure}
\subsubsection{\textbf{Shared memory requests are intensive }}
If the shared memory access is intensive, its latency can contribute to the final execution time significantly. Fig. \ref{fig:matrixMul_ShMem} shows matrix multiplication with shared memory as an example. At the beginning all the warps load data from global memory. Then in the second phase each warp access the shared memory for multiple times, which consumes much time. Since the total shared memory latency of phase 2 is longer enough to hide global memory latency of other blocks, the global requests within the same block have no contention with others. Due to the function of synchronization, we can regard this procedure as repetition of these two big steps. Thus, the total execution time of one round can be calculated by Eq. \eqref{s1_matmul}$\sim$\eqref{eq:t_matmul}. $i\_itrs$ denotes the number of shared memory transactions within $phase2$.
\begin{equation} \label{s1_matmul}
\begin{aligned}
T_{phase1} = &avr\_comp \times 2 + agl\_del \times gld\_trans \\
&\times \#Aw + agl\_lat + sh\_lat \\
\end{aligned}
\end{equation}
\begin{equation} \label{s2_matmul}
\begin{aligned}
T_{phase2} = &avr\_comp \times (warps\_per\_block - 1) 	\\
&+(avr\_comp + sh\_lat) \times i\_itrs 			\\
\end{aligned}
\end{equation}
\begin{equation} \label{s3_matmul}
\begin{aligned}
T_{phase3} = & avr\_comp \times 2 + agl\_del \times gld\_trans \\
& \times \#Wpb + agl\_lat + sh\_lat	\\
\end{aligned}
\end{equation}
\begin{equation} \label{eq:t_matmul}
\begin{split}
T_{active} = &T_{phase1} + (T_{phase2} + T_{phase3}) \times o\_itrs	\\
\end{split}
\end{equation}

\begin{table*}[ht]
	\centering
	\caption{Parameters used for Performance Modeling}
	\label{tab:paraModel}
	\scriptsize{
		\begin{tabular}{|p{0.5in}|p{3.3in}|p{1.7in}|} \hline
			\textbf{Parameters} & \textbf{Definition} & \textbf{How to achieve}	\\ \hline
			agl\_lat & average latency of global memory transaction considering L2 cache hit rate & Equation \eqref{eq:agl_lat_a}\\ \hline		
			agl\_del &  average delay of global memory transaction considering L2 cache hit rate & Equation \eqref{eq:agl_lat_a}\\ \hline	
			dm\_lat & dram latency of one global transaction & microbenchmarking\\ \hline		
			dm\_del & dram delay of one global transaction & microbenchmarking\\ \hline	
			interArr &  Inter-arrival time between two consequent memory requests  & microbenchmarking\\ \hline
			l2\_lat & L2 cache latency of one global transaction & microbenchmarking\\ \hline		
			l2\_del &  L2 cache delay of one global transaction & hardware specification\\ \hline	
			l2\_hr  &  Hit rate at L2 cache for all transactions from SM  & Nsight profiling\\ \hline
			sh\_lat & latency of one shared memory transaction & microbenchmarking \\ \hline			
			gld\_trans & Number of global load/store transactions of one warp in one iteration & Nsight profiling\\ \hline
			comp\_inst  & total compute instructions of the kernel  & Nsight profiling\\ \hline
			avr\_comp  & average computation time of one period   & Equation \eqref{eq:av_com_b}\\ \hline
			inst\_cycle  & latency for each computation instruction  & hardware specification \\ \hline
			\#B	 & Total number of blocks & kernel setup\\ \hline		
			\#Wpb & Number of warps per block & kernel setup\\ \hline	
			\#W	& Number of total warps of a kernel & kernel setup\\ \hline	
			\#Asm  	& 	Active number of SMs 	& Nsight profiling	\\ \hline
			\#Aw	  & Number of warps run concurrently on one SM & Nsight profiling \\ \hline
			o\_itrs  & Number of first level iteration within a thread & source code analysis \\ \hline
			i\_itrs  & Number of second level iteration within a thread &  source code analysis \\ \hline	
			$T_{lat}$   & Total execution time of multiple global memory requests & Equation \eqref{eq:total_lat_not_full}, \eqref{eq:total_lat_sat} \\  \hline
			$T_{active}$   & Cycles for executing one round of active warps on a SM & Equation \eqref{eq:comp-case-cycle}, \eqref{eq:mem-case-cycle},  \eqref{eq:new1-case-cycle}, \eqref{eq:new2-case-cycle}, \eqref{eq:shm-cycle-if}, \eqref{eq:t_matmul} \\ \hline
			$T_{exec}$   & Total execution time of a kernel & Equation \eqref{eq:time_exec} \\  \hline
			$core\_f$   & frequency that controls the speed of SM & Adjustments\\ \hline
			$mem\_f$   & frequency that controls the speed of DRAM &  Adjustments\\ \hline
		\end{tabular}
	}
\end{table*}

Matrix multiplication with shared memory optimization is one instance of this case. Its performance is sensitive to both core and memory frequency which is revealed in our previous motivating examples and it can be explained by our model. First, $T_{phase1}$ and $T_{phase3}$ contain a large number of global memory transactions which makes the execution time sensitive to memory frequency. Second, although shared memory latency is much shorter than global memory latency, nearly 3 dozens of shared memory requests in $T_{phase2}$ also contribute a lot to the final $T_{active}$, which makes the execution time sensitive to core frequency as well.
These two cases can be adopted to most classical GPU kernels. Though there are other more complicated irregular instances such as MC\_EstimatePiInlineP and reduction, the similar methodology of phase partition can somehow apply to them. We leave detailed analysis for these irregular kernels for future work.

\section{Experiments} \label{sec:Ex}
\subsection{Experimental Methodology}
With the help of NVIDIA Inspector \cite{nvidiaInspector}, we can fix the performance state and adjust core frequency and memory frequency of GPU together within a certain range. By this method we can obtain execution time data with certain frequency combinations of GPU. We cover both core frequency and memory frequency at a 2.5x range of scaling from 400 MHz to 1000 MHz with a step size of 100 MHz so that totally 49 frequency combinations are tested.

\begin{table}[ht]
	\centering
	\caption{Target GPU frequency configurations}
	\label{tab:VFconfiguration}
	\scriptsize{
		\begin{tabular}{|p{1.2in}|p{1in}|} \hline
			\textbf{Device} & \textbf{GTX 980}	\\ \hline
			Compute apability 		& 5.2			\\ \hline
			SMs * cores per SM 	   	& 16 * 128				\\  \hline
			Global Memory bus width & 256-bit		\\ \hline
			Global Memory size      & 4GB			\\ \hline
			\hline
			Core frequency scaling & [400MHz, 1000MHz]	\\ \hline
			Memory scaling & [400MHz, 1000MHz]     	\\ \hline
			Scaling stride & 100MHz					\\ \hline
			
		\end{tabular}
	}
\end{table}

We use NVIDIA Nsight tools \cite{NsightVS} to extract the performance counters we need to drive our model at the baseline frequency 700 MHz for both core and memory. Note that we only need one time data collection with this method, which makes our model work fast. We choose 700 MHz for baseline since it leaves the space for raising and declining the frequency, which suggests that our performance model is more general and flexible. 

\begin{table}[ht]
	\centering
	\caption{Tested Applications}
	\label{tab:testedApp}
	\scriptsize{
		\begin{tabular}{|p{0.9in}|p{1.3in}|} \hline
			\textbf{abbr.} & \textbf{Application Name}	\\ \hline
			BS 		& BlackScholes		\\ \hline
			CG   &  conjugateGradient  \\ \hline
			FWT     & fastWalshTransform \\ \hline
			MMG 	   	& matrixMul(Global)				\\  \hline
			MMS & matrixMul(Shared)		\\ \hline
			SC   & scan 		\\ \hline
			SN   & sortingNetworks  \\ \hline
			SP    & scalarProd  \\ \hline
			TR      & transpose			\\ \hline
			VA     & vector addition			\\ \hline
			convSp  & convolutionSeparable \\ \hline
			
		\end{tabular}
	}
\end{table}

We validate our model among 12 realistic GPU kernels from CUDA SDK 6.5 listed in Table \ref{tab:testedApp} on a real NVIDIA Maxwell GTX980. Hardware specifications of our test machine are listed in Table \ref{tab:VFconfiguration}. These benchmark applications cover a wide range of execution pattern such as DRAM intensive, L2 cache intensive, shared memory intensive and computation intensive. We repeat our experiments for 1000 times and report the average results.

\begin{figure}[ht]
	\centering
	\includegraphics[width=0.5\textwidth]{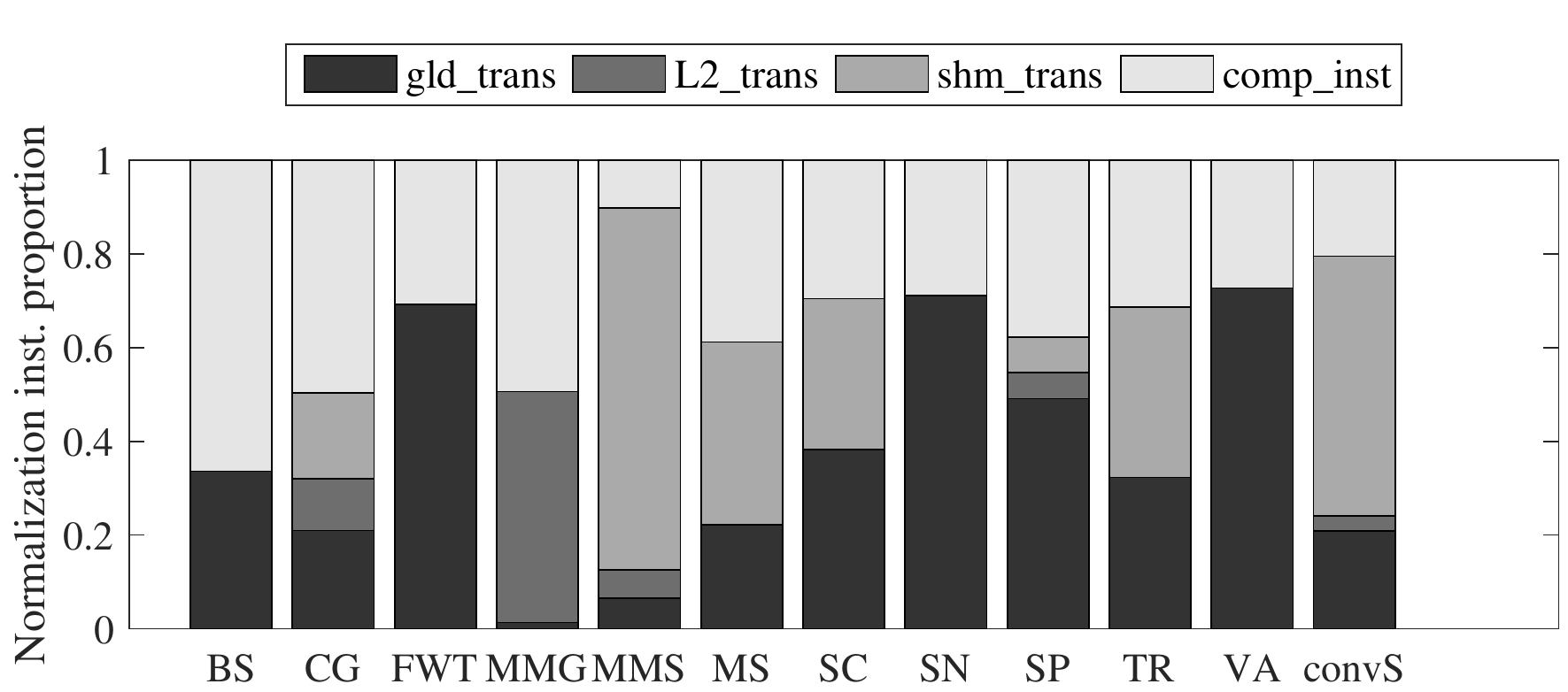}\\
	\caption{Breakdown of different types of instructions}
	\label{fig:inst_bd}
\end{figure}

\begin{figure*}[ht]
	\centering     
	\subfigure
	{
		\includegraphics[width=1\linewidth]{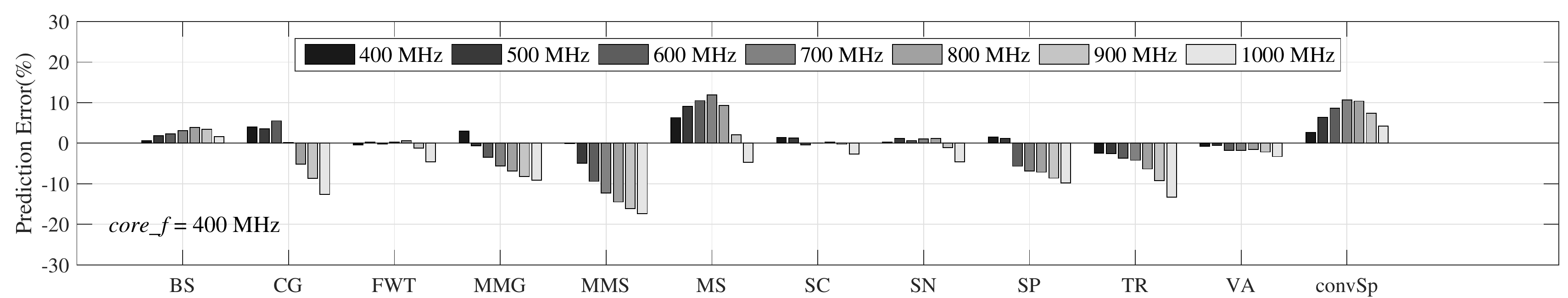}
		\label{fig:a}
	}
	\subfigure
	{
		\includegraphics[width=1\linewidth]{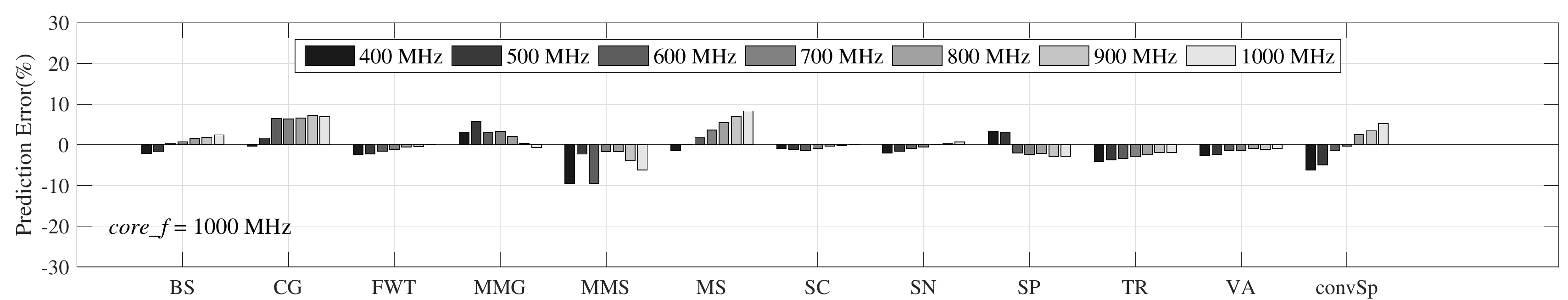}
		\label{fig:b}
	}
	\subfigure
	{
		\includegraphics[width=1\linewidth]{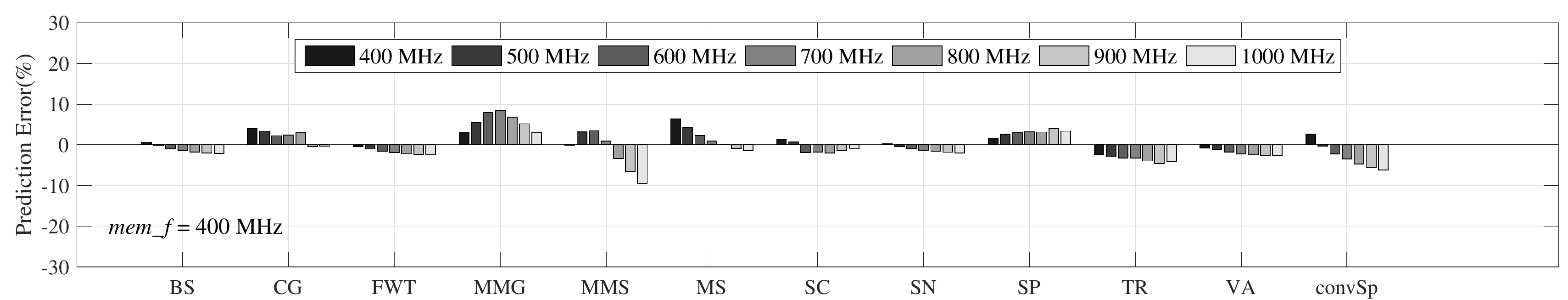}
		\label{fig:c}
	}
	\subfigure
	{
		\includegraphics[width=1\linewidth]{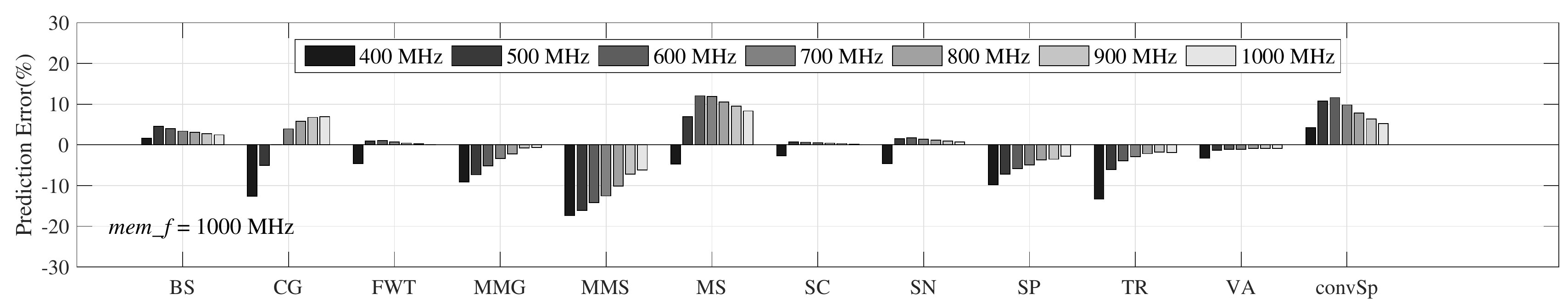}
		\label{fig:d}
	}
	\caption{\label{fig:predErr} Time Prediction Error under different frequency settings. Each figure shows the results of scaling one of the frequencies when the other is fixed.}
	
\end{figure*}

\subsection{Experimental Results}
First, we would like to observe the instruction distributions of different GPU kernels with the help of NVIDIA Nsight tools. As Fig. \ref{fig:inst_bd} demonstrates, our tested kernels have various partitions of different types of instructions, which suggests that we attempt to design a general model for different types of GPU kernels. In addition, such instruction statistics help us locate the principle contributors to the execution time under certain frequency settings. Combined with experimental results, we can also infer some error resources of under- or over-estimation of execution time. 
\begin{figure}[ht]
	\centering
	\includegraphics[width=0.5\textwidth]{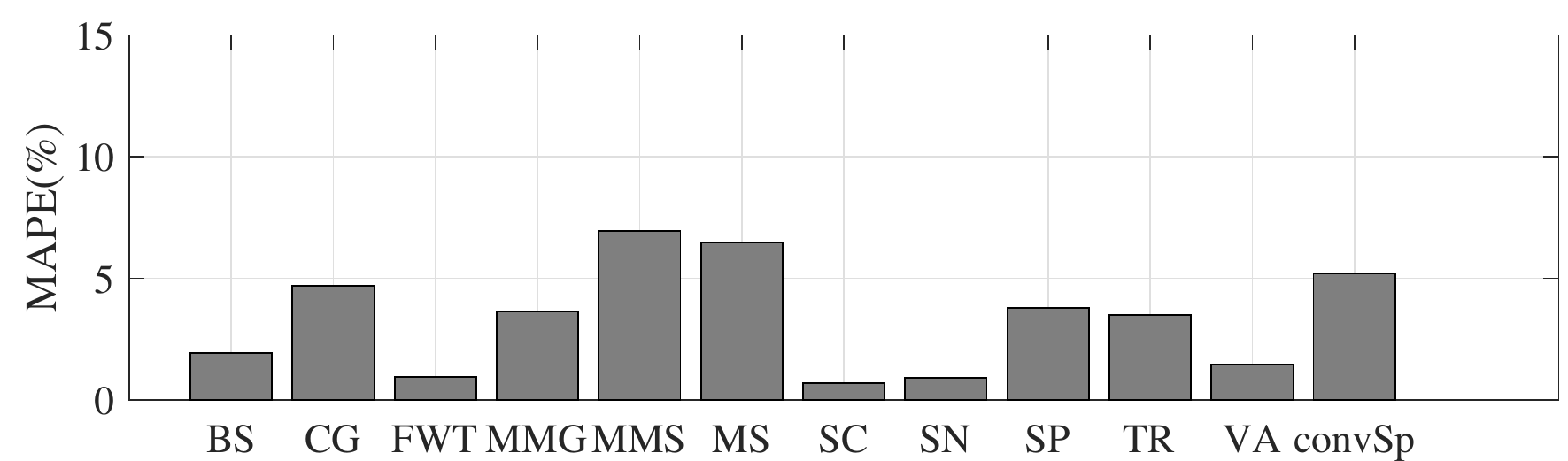}\\
	\caption{Mean absolute percentage error average across all available frequency pairs}
	\label{fig:MAPE_overview}
\end{figure}

Fig. \ref{fig:predErr} shows the time prediction error under varing memory frequency with fixed core frequency and vice versa. Across 49 available frequency settings among 12 kernels, we achieve below 16\% error for each prediction and even 90\% of them are under 10\%. As for each kernels, the mean absolute percentage error (MAPE) ranges from 0.7\% to 6.9\% as shown in Fig. \ref{fig:MAPE_overview}. We achieve 3.5\% MAPE across all the testing samples. As mentioned before, some prediction errors can be explained by instruction distributions in a kernel. For example, MatMul(S) have relative bigger under-estimation errors in \ref{fig:a} than those in \ref{fig:b}. The possible reason is that the time consumption in SM, shared memory access in particular, is under-estimated. As for MatMul(G), although it launches a great number of global memory transactions, it has a high L2 cache hit rate up to 97.5\%, which results in its sensitivity to core frequency as well. Some kernels like convSp, FWT and SP have approximately linear decreasing errors with larger memory frequency in \ref{fig:a} and \ref{fig:b} but stable errors in \ref{fig:c} and \ref{fig:d}. These also reveal that this kind of kernels are more sensitive to memory frequency, which are also supported by the facts that they have high proportion of DRAM transactions. 

\section{Conclusion} \label{sec:cc}
In this work, we demonstrated a GPGPU performance predictor for a wide range of both core and memory frequency. We first estimate the total time consumption of multiple memory requests under different frequency settings. Then our predictor takes the profiling data of a given kernel under our baseline frequency setting as input to estimate the performance of it at other core and memory frequencies. Our model can predict the execution time of a GPU kernel on real hardware quickly and accurately, which is important to derive real-time energy conservation suggestions with DVFS techniques. 

We shows that our performance estimation method can achieve 3.8\% MAPE across up to 2.5x both core and memory frequency scaling. Our experimental results also indicate that our model can catch not only the performance scaling behaviors of DRAM very precisely but also L2 cache and shared memory. 

As for future work, we have two directions of improvements. First, our model does not explore too much about shared memory as we treat on DRAM and even does not take texture/L1 cache and constant memory into account, which may introduce larger error for kernels containing access requests to them. Second, collaborated with GPU power models, it is potentially a remarkable project to build a real-time voltage and frequency controller for GPU based on energy conservations strategies with DVFS techniques.


\section*{Acknowledgment}
This work is supported by Shenzhen Basic Research Grant SCI-2015-SZTIC-002.



%

\bibliographystyle{IEEEtran}
\bibliography{GPGPU Performance Estimation with Core and Memory Frequency Scaling.bbl}

\begin{thebibliography}{10}
\providecommand{\url}[1]{#1}
\csname url@samestyle\endcsname
\providecommand{\newblock}{\relax}
\providecommand{\bibinfo}[2]{#2}
\providecommand{\BIBentrySTDinterwordspacing}{\spaceskip=0pt\relax}
\providecommand{\BIBentryALTinterwordstretchfactor}{4}
\providecommand{\BIBentryALTinterwordspacing}{\spaceskip=\fontdimen2\font plus
\BIBentryALTinterwordstretchfactor\fontdimen3\font minus
  \fontdimen4\font\relax}
\providecommand{\BIBforeignlanguage}[2]{{%
\expandafter\ifx\csname l@#1\endcsname\relax
\typeout{** WARNING: IEEEtran.bst: No hyphenation pattern has been}%
\typeout{** loaded for the language `#1'. Using the pattern for}%
\typeout{** the default language instead.}%
\else
\language=\csname l@#1\endcsname
\fi
#2}}
\providecommand{\BIBdecl}{\relax}
\BIBdecl

\bibitem{collobert2011torch7}
R.~Collobert, K.~Kavukcuoglu, and C.~Farabet, ``{Torch7: A matlab-like
  environment for machine learning},'' in \emph{BigLearn, NIPS Workshop}, no.
  EPFL-CONF-192376, 2011.

\bibitem{abadi2015tensorflow}
M.~Abadi, A.~Agarwal, P.~Barham, E.~Brevdo, Z.~Chen, C.~Citro, G.~S. Corrado,
  A.~Davis, J.~Dean, M.~Devin \emph{et~al.}, ``{TensorFlow: Large-scale machine
  learning on heterogeneous systems, 2015},'' \emph{Software available from
  tensorflow. org}, vol.~1, 2015.

\bibitem{jia2014caffe}
Y.~Jia, E.~Shelhamer, J.~Donahue, S.~Karayev, J.~Long, R.~Girshick,
  S.~Guadarrama, and T.~Darrell, ``{Caffe: Convolutional architecture for fast
  feature embedding},'' in \emph{Proceedings of the 22nd ACM international
  conference on Multimedia}, 2014, pp. 675--678.

\bibitem{cntkspeedcomparasion2016}
X.~Huang, ``{Microsoft Computational Network Toolkit offers most efficient
  distributed deep learning computational performance},''
  \url{https://goo.gl/9UUwVn}, 2015, accessed: 2016-07-12.

\bibitem{TitanIntro}
O.~R.~N. Laboratory, ``{Introducing {Titan}: advancing the era of accelerating
  computing},'' [Online] https://www.olcf.ornl.gov/titan/.

\bibitem{mei2016survey}
\BIBentryALTinterwordspacing
X.~Mei, Q.~Wang, and X.~Chu, ``{A Survey and Measurement Study of GPU DVFS on
  Energy Conservation},'' \emph{Accepted by Digital Communication and Network
  (DCN)}. [Online]. Available: \url{https://arxiv.org/abs/1610.01784}
\BIBentrySTDinterwordspacing

\bibitem{bridges2016understanding}
R.~A. Bridges, N.~Imam, and T.~M. Mintz, ``Understanding gpu power: A survey of
  profiling, modeling, and simulation methods,'' \emph{ACM Computing Surveys
  (CSUR)}, vol.~49, no.~3, p.~41, 2016.

\bibitem{greenmetrics2017power}
\BIBentryALTinterwordspacing
Q.~Wang and X.~Chu, ``{GPGPU Power Estimation with Core and Memory Frequency
  Scaling},'' \emph{SIGMETRICS Perform. Eval. Rev.}, vol.~45, no.~2, pp.
  73--78, Oct. 2017. [Online]. Available:
  \url{http://doi.acm.org/10.1145/3152042.3152066}
\BIBentrySTDinterwordspacing

\bibitem{hpca2018power}
J.~Guerreiro, A.~Ilic, N.~Roma, and P.~Tomas, ``{GPGPU Power Modeling for
  Multi-domain Voltage-Frequency Scaling},'' in \emph{2018 IEEE International
  Symposium on High Performance Computer Architecture (HPCA)}, Feb 2018, pp.
  789--800.

\bibitem{hong2009analytical}
S.~Hong and H.~Kim, ``{An Analytical Model for a GPU Architecture with
  Memory-level and Thread-level Parallelism Awareness},'' in \emph{Proceedings
  of the 36th Annual International Symposium on Computer Architecture}, ser.
  ISCA '09.\hskip 1em plus 0.5em minus 0.4em\relax New York, NY, USA: ACM,
  2009, pp. 152--163.

\bibitem{hong2010integrated}
------, ``{An Integrated GPU Power and Performance Model},'' in
  \emph{Proceedings of the 37th Annual International Symposium on Computer
  Architecture}, ser. ISCA '10.\hskip 1em plus 0.5em minus 0.4em\relax New
  York, NY, USA: ACM, 2010, pp. 280--289.

\bibitem{nath2015crisp}
R.~Nath and D.~Tullsen, ``{The CRISP performance model for dynamic voltage and
  frequency scaling in a GPGPU},'' in \emph{Proceedings of the 48th
  International Symposium on Microarchitecture}.\hskip 1em plus 0.5em minus
  0.4em\relax ACM, 2015, pp. 281--293.

\bibitem{miftakhutdinov2012predicting}
R.~Miftakhutdinov, E.~Ebrahimi, and Y.~N. Patt, ``Predicting performance impact
  of dvfs for realistic memory systems,'' in \emph{Proceedings of the 2012 45th
  Annual IEEE/ACM International Symposium on Microarchitecture}.\hskip 1em plus
  0.5em minus 0.4em\relax IEEE Computer Society, 2012, pp. 155--165.

\bibitem{sim2012performance}
J.~Sim, A.~Dasgupta, H.~Kim, and R.~Vuduc, ``{A performance analysis framework
  for identifying potential benefits in GPGPU applications},'' in \emph{ACM
  SIGPLAN Notices}, vol.~47, no.~8.\hskip 1em plus 0.5em minus 0.4em\relax ACM,
  2012, pp. 11--22.

\bibitem{chen2014run}
X.~Chen, Y.~Wang, Y.~Liang, Y.~Xie, and H.~Yang, ``Run-time technique for
  simultaneous aging and power optimization in gpgpus,'' in \emph{Design
  Automation Conference (DAC), 2014 51st ACM/EDAC/IEEE}.\hskip 1em plus 0.5em
  minus 0.4em\relax IEEE, 2014, pp. 1--6.

\bibitem{leng2013gpuwattch}
J.~Leng, T.~Hetherington, A.~ElTantawy, S.~Gilani, N.~S. Kim, T.~M. Aamodt, and
  V.~J. Reddi, ``Gpuwattch: Enabling energy optimizations in gpgpus,'' in
  \emph{Proceedings of the 40th Annual International Symposium on Computer
  Architecture}, ser. ISCA '13.\hskip 1em plus 0.5em minus 0.4em\relax New
  York, NY, USA: ACM, 2013, pp. 487--498.

\bibitem{lucas2013single}
J.~Lucas, S.~Lal, M.~Andersch, M.~Alvarez-Mesa, and B.~Juurlink, ``{How a
  single chip causes massive power bills GPUSimPow: A GPGPU power simulator},''
  in \emph{Performance Analysis of Systems and Software (ISPASS), 2013 IEEE
  International Symposium on}.\hskip 1em plus 0.5em minus 0.4em\relax IEEE,
  2013, pp. 97--106.

\bibitem{bakhoda2009analyzing}
A.~Bakhoda, G.~L. Yuan, W.~W. Fung, H.~Wong, and T.~M. Aamodt, ``{Analyzing
  CUDA workloads using a detailed GPU simulator},'' in \emph{Performance
  Analysis of Systems and Software, 2009. ISPASS 2009. IEEE International
  Symposium on}.\hskip 1em plus 0.5em minus 0.4em\relax IEEE, 2009, pp.
  163--174.

\bibitem{aamodt2012gpgpu}
T.~M. Aamodt, W.~W. Fung, I.~Singh, A.~El-Shafiey, J.~Kwa, T.~Hetherington,
  A.~Gubran, A.~Boktor, T.~Rogers, A.~Bakhoda \emph{et~al.}, ``{GPGPU-Sim 3. x
  manual},'' 2012.

\bibitem{wu2015gpgpu}
G.~Wu, J.~L. Greathouse, A.~Lyashevsky, N.~Jayasena, and D.~Chiou, ``{GPGPU
  performance and power estimation using machine learning},'' in \emph{High
  Performance Computer Architecture (HPCA), 2015 IEEE 21st International
  Symposium on}.\hskip 1em plus 0.5em minus 0.4em\relax IEEE, 2015, pp.
  564--576.

\bibitem{abe2014power}
Y.~Abe, H.~Sasaki, S.~Kato, K.~Inoue, M.~Edahiro, and M.~Peres, ``{Power and
  performance characterization and modeling of gpu-accelerated systems},'' in
  \emph{Parallel and Distributed Processing Symposium, 2014 IEEE 28th
  International}.\hskip 1em plus 0.5em minus 0.4em\relax IEEE, 2014, pp.
  113--122.

\bibitem{song2013simplified}
S.~Song, C.~Su, B.~Rountree, and K.~W. Cameron, ``{A simplified and accurate
  model of power-performance efficiency on emergent gpu architectures},'' in
  \emph{Parallel \& Distributed Processing (IPDPS), 2013 IEEE 27th
  International Symposium on}.\hskip 1em plus 0.5em minus 0.4em\relax IEEE,
  2013, pp. 673--686.

\bibitem{nagasaka2010statistical}
H.~Nagasaka, N.~Maruyama, A.~Nukada, T.~Endo, and S.~Matsuoka, ``{Statistical
  power modeling of GPU kernels using performance counters},'' in \emph{Green
  Computing Conference, 2010 International}.\hskip 1em plus 0.5em minus
  0.4em\relax IEEE, 2010, pp. 115--122.

\bibitem{dao2015performance}
T.~T. Dao, J.~Kim, S.~Seo, B.~Egger, and J.~Lee, ``{A performance model for
  gpus with caches},'' \emph{Parallel and Distributed Systems, IEEE
  Transactions on}, vol.~26, no.~7, pp. 1800--1813, 2015.

\bibitem{cudaProgram}
NVIDIA, ``{CUDA C Programming Guide},'' [Online]
  http://docs.nvidia.com/cuda/cuda-c-programming-guide/index.html.

\bibitem{kursun2006supply}
V.~Kursun and E.~G. Friedman, ``{Supply and Threshold Voltage Scaling
  Techniques},'' \emph{Multi-Voltage CMOS Circuit Design}, pp. 45--84, 2006.

\bibitem{kim2015racing}
D.~H. Kim, C.~Imes, and H.~Hoffmann, ``{Racing and Pacing to Idle: Theoretical
  and Empirical Analysis of Energy Optimization Heuristics},'' in
  \emph{Cyber-Physical Systems, Networks, and Applications (CPSNA), 2015 IEEE
  3rd International Conference on}.\hskip 1em plus 0.5em minus 0.4em\relax
  IEEE, 2015, pp. 78--85.

\bibitem{wong2010demystifying}
H.~Wong, M.-M. Papadopoulou, M.~Sadooghi-Alvandi, and A.~Moshovos,
  ``{Demystifying GPU microarchitecture through microbenchmarking},'' in
  \emph{Performance Analysis of Systems \& Software (ISPASS), 2010 IEEE
  International Symposium on}.\hskip 1em plus 0.5em minus 0.4em\relax IEEE,
  2010, pp. 235--246.

\bibitem{meltzer2013micro}
R.~Meltzer, C.~Zeng, and C.~Cecka, ``{Micro-benchmarking the C2070},'' in
  \emph{GPU Technology Conference}.\hskip 1em plus 0.5em minus 0.4em\relax
  Citeseer, 2013.

\bibitem{mei2014npc}
X.~Mei, K.~Zhao, C.~Liu, and X.~Chu, ``{"Benchmarking the Memory Hierarchy of
  Modern GPUs"},'' in \emph{Network and Parallel Computing}, 2014, pp.
  144--156.

\bibitem{mei2015tpds}
X.~Mei and X.~Chu, ``{Dissecting GPU Memory Hierarchy through
  Microbenchmarking},'' \emph{IEEE Transactions on Parallel and Distributed
  Systems}, doi:10.1109/TPDS.2016.2549523.

\bibitem{hennessy2011computer}
J.~L. Hennessy and D.~A. Patterson, \emph{{Computer architecture: a
  quantitative approach}}.\hskip 1em plus 0.5em minus 0.4em\relax Elsevier,
  2011.

\bibitem{nvidiaInspector}
Orbmu2k, ``{NVIDIA Inspector},'' [Online]
  http://blog.orbmu2k.de/tools/nvidia-inspector-tool.

\bibitem{NsightVS}
NVIDIA, ``{NVIDIA Nsight Visual Studio Edition},'' [Online]
  https://developer.nvidia.com/nvidia-nsight-visual-studio-edition.

\end{thebibliography}

\end{document}